%% file: main.tex
\documentclass[a4paper,11pt]{article}

\usepackage{jheppub} 

\usepackage{xcolor}
\usepackage{longtable,array}
\usepackage{tikz}
\usetikzlibrary{angles,quotes,arrows.meta,calc}

\usepackage[T1]{fontenc} 

\usepackage{dsfont,calc,bbm,accents,leftindex,slashed}

\def\beq{\begin{equation}}
\def\eeq{\end{equation}}

\def\bes{\begin{subequations}}
\def\ees{\end{subequations}}

\def\d{\mathbbm{d}}

\def\D{\textsc{D}}

\def\P{\textsc{S}}

\def\u{{{\Upsilon}}}
\def\s{\Sigma}
\def\i{\textsc{I}}

\newcommand{\In}[1]{\hspace{-#1pt}\int\hspace{-#1pt}}
\newcommand{\BIn}[3]{%
  \hspace{-1pt * \real{#1}}
  \int_{#2}^{#3}%
  \hspace{-1pt * \real{#1}-9pt}%
}

\newcommand{\3}[1]{{\boldsymbol{#1}}}

\newcommand{\0}[1]{\accentset{\scalebox{0.5}{$\circ$}}{#1}}

\newcommand{\1}[1]{\accentset{\scalebox{0.5}{$\bullet$}}{#1}}

\newcommand{\di}[1]{\accentset{\raisebox{-0.2ex}[0.2ex][0pt]{\scalebox{0.6}{$\boldsymbol\sim$}}}{#1}}

\newcommand{\zu}[1]{\accentset{\scalebox{0.5}{$\circ(\bullet)$}}{#1}}

\newcommand{\al}[1]{\begin{align}{#1}\end{align}}

\definecolor{Ycolor}{RGB}{23,86,126}
\definecolor{Sigmacolor}{RGB}{235,126,110}
\definecolor{Icolor}{RGB}{33,132,161}

\preprint{UMN-TH-4602/26}

\title{\boldmath 
Unstable particle collisions in space and time
}

\author[a,b]{Irene Corral Lorences,}
\author[c]{Zhen Liu,}
\author[d,e]{Giuliano Panico,}
\author[f,g]{and Andrea Wulzer}

\affiliation[a]{Freie Universität Berlin, Berlin, Germany}
\affiliation[b]{Departament de F\'isica, Universitat Aut\`onoma de Barcelona, 08193 Bellaterra, Barcelona, Spain}
\affiliation[c]{School of Physics and Astronomy, University of Minnesota, Minneapolis, MN 55455, USA}
\affiliation[d]{Dipartimento di Fisica e Astronomia, Universit\`a degli Studi di Firenze, \protect\\ Via G. Sansone 1, 50019 Sesto Fiorentino, Italy} 
\affiliation[e]{INFN, Sezione di Firenze, Via G. Sansone 1, 50019 Sesto Fiorentino, Italy}
\affiliation[f]{Institut de F\'{\i}sica d'Altes Energies (IFAE), The Barcelona Institute of Science and Technology (BIST),
Campus UAB, 08193 Bellaterra, Barcelona, Spain}
\affiliation[g]{ICREA, Instituci\'o Catalana de Recerca i Estudis Avan\c{c}ats, 
Passeig de Llu\'{\i}s Companys 23, 
08010 Barcelona, Spain}

\emailAdd{irene.corral@fu-berlin.de}
\emailAdd{zliuphys@umn.edu}
\emailAdd{giuliano.panico@unifi.it}
\emailAdd{awulzer@ifae.es}

\abstract{
A proper quantum description of colliding beams reveals the spacetime structure of collision processes, going beyond the conventional picture of classical point-like particles undergoing short-distance scattering. This familiar picture emerges as an approximation in the classical beam limit and under specific conditions. Resolving the $t$-channel singularity in processes initiated by an unstable particle requires going beyond the short-distance treatment because this singularity is associated with a long-distance effect: the decay of the unstable particle, followed by the propagation of its decay products until they scatter with the other beam.\\[3pt]
\noindent{}In this article, we develop the theory of spacetime collisions to allow the systematic study of the classical expansion accessing quantum corrections, and we apply it to unstable-particle collisions. At all orders in the expansion, we find that long-distance effects associated with unstable-particle decay coexist with direct short-distance interactions with the other beam. At leading order, the short-distance component is described by a cross section. This allows us to define and compute the unstable-particle scattering cross section through a uniquely specified regularization prescription for the $t$-channel singularity. Prospective applications include cross-section calculations for muon experiments, including a muon collider. It would also be interesting to design experiments to observe the novel effects arising from the quantum nature of the colliding beams that emerge in our formalism.}

\usepackage[compat=1.1.0]{tikz-feynman}
\begin{document}

\maketitle
\flushbottom

\section{Introduction}

\begin{figure}[t]
\centering
\includegraphics[width=\textwidth]{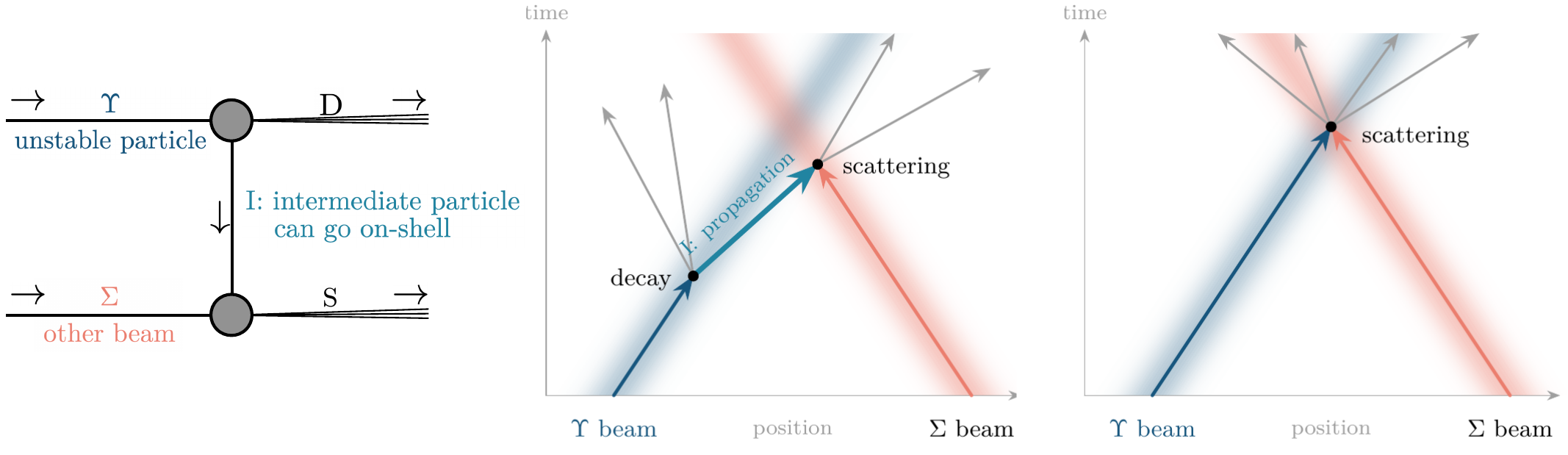}
\caption{{\bf{Left:}} Feynman diagram of the $\u\,\s\to\D\P$ process. {\bf{Middle:}} spacetime representation of the $\u$ particle decay, the propagation and finally the scattering of its decay product $\i$. {\bf{Right:}} direct $\u$-$\s$ short-distance interaction.}\label{fig:one}
\end{figure}

The regular cross section calculation is ill-defined for a large class of processes initiated by an unstable initial particle $\u$. The problematic processes are those that combine---see the left panel of figure~\ref{fig:one}---a diagram describing an allowed decay $\u\to\i\,\D$ of the particle $\u$ and a diagram describing an allowed scattering $\i\,\s\to\P$ of the other initial state $\s$ with the intermediate particle $\i$ produced in the $\u$ decay. Both the decay and the scattering kinematics can be satisfied with an on-shell particle $\i$. Therefore, there exists a point in the accessible phase space at which the $\i$ momentum goes on shell producing a $t$-channel singularity. Since the singularity is non-integrable, the cross section calculation is ill-defined.

%
%

This problem has been known since a very long time~\cite{Peierls:1961zz} and it has been investigated in refs.~\cite{Ginzburg:1995js,Melnikov:1996na,Melnikov:1996iu,Melnikov:1996ft,Dams:2002uy,Giacosa:2025fvq,Asai:2025cdb,Iglicki:2022jjf}. Particular attention has been devoted to the process $\mu^-\mu^+\to e^-\bar\nu_e W^+$ at a future high-energy muon collider, but other applications have also been considered. This literature clarified~\cite{Melnikov:1996na,Melnikov:1996iu,Melnikov:1996ft} that the singularity is regulated by the finite extension in space of the colliding beams and studied the effect of the regulated singularity on the transition probability in the large-beam limit. The result has a simple classical interpretation schematized in the middle panel of figure~\ref{fig:one}. It corresponds to the effect of the decay of particle $\u$ followed by the propagation of its decay product $\i$ over a macroscopic distance until it reaches and scatters with a $\s$ particle in the second beam. Equivalently, the regulated singularity corresponds to the formation of a beam of $\i$ particles from the $\u$ decay---specifically, a $\nu_\mu$ beam in the case of the muon collider process---and its interaction with the other beam. The possible role played by the finite width of $\u$ has been also emphasised~\cite{Ginzburg:1995js,Dams:2002uy}. While not connected with the singularity resolution (namely, the singularity is resolved even in the zero-width limit) finite-width effects should be taken into account for a complete treatment of the problem. We elaborate on this aspect in the Conclusions. 

While they clarify the physical origin and the resolution of the divergence in the regular cross section calculation, these results do not address the original question of how the unstable particle cross section should be computed instead. The classical process of $\u$ decay followed by $\i$ propagation and eventually scattering, which emerges at the leading order in the large-beam limit, involves the cross section for the $\i$ particle and not for the $\u$ particle scattering. The $\u$ particle cross section must emerge at higher order after the subtraction of the regulated singularity. It is expectedly obtained by supplementing the regular cross section calculation with a prescription to deal with the $t$-channel singularity. The proposal of refs.~\cite{Melnikov:1996na,Melnikov:1996iu,Melnikov:1996ft} was to exclude a finite region around the singularity in the phase-space integration, with the width of the excluded region serving as a conventional regulator. An alternative was proposed in~\cite{Asai:2025cdb}. Establishing a principled definition of the unstable particle cross section and deriving the corresponding rules for computing it is one of the motivations of the present work. 

The technical foundations of our study were laid down in ref.\,\cite{Kotkin:1992bj} (see also refs.\,\cite{Ginzburg:1991mv,Ginzburg:1991mw} and \cite{Volobuev:2019zan}) with the purpose of calculating finite beam size effects on electromagnetic processes at colliders. The approach is based on the proper quantum description of the colliding beams and on the direct calculation of the transition probability without the intermediate notion of a ``cross section''. This is essential for our study because the cross section offers an adequate description of the physical collision process only when it can be schematized as the short-distance (i.e., exactly localized in space and instantaneous in time, as depicted on the right panel of figure~\ref{fig:one}) interaction of point-like particles. The process initiated by the decay of the $\u$ particle (middle panel) entails instead the propagation of the $\i$ particle by a finite distance and therefore it is a long-distance effect. We thus need a formalism where the spacetime structure of the collision process emerges from the calculation and the short-distance nature of the collision is not assumed as in the regular approach to collision theory.

In section\,\ref{sec:2} we review but also improve the spacetime collision theory formalism. Significant innovations are the definition of a Lorentz-scalar (see also appendix~\ref{sec:relwd}) Wigner density and of the classical beam limit in which measurements---specifically, measurements of conserved currents, see also appendix~\ref{app:LM}---performed on the beam produce classical results. The end result is that the transition probability is the convolution of the beam density matrices---or Wigner densities, in coordinate space---with a ``collision function'' that carries all the information on the microscopic process and plays a similar role to the cross section in the regular formalism. The collision function depends on a momentum variable $\di{p}$ and the classical limit is when $\di{p}$ is small. The terms in the collision function that are analytic at $\di{p}=0$ produce short-distance contributions to the transition probability.

Unstable-particle collision is studied in section\,\ref{sec:3}. The $t$-channel singularity in the regular cross section calculation corresponds to the emergence of a non-analytic term in the collision function that coexists with an analytic one. The two contributions can be disentangled and studied separately. At all orders in the classical expansion the contribution to the transition probability of the non-analytic term has a long-distance structure that corresponds to the decay of $\u$, the propagation and finally the scattering of $\i$. At the leading order in the expansion the result matches the one of the classical calculation. The contribution of the analytic term is instead of the short-distance type and at the leading order in the classical expansion it is described by a cross section. This provides us with the definition of the unstable particle cross section and, in turn, with the rules for computing it. The above results are established for generic processes and cross-checked by an explicit calculation in section\,\ref{sec:exc} for a simple $2\to2$ collision. The explicit calculation reveals an interesting enhancement of the quantum corrections near to the threshold for on-shell exchange. Appendix\,\ref{sec:dist} reviews results on the series expansion of distributions that are used in the derivation. Appendix\,\ref{app:notation} offers a summary of the notation employed throughout the paper.

Section~\ref{sec:conc} reports our conclusions and an outlook to future work. This includes cross section calculations at current and planned experiments that collide unstable particles, but possibly also the design of experiments dedicated to the detection of the unstable particle cross section and even more exotic quantum effects in the transition probability that cannot be described by a classical probabilistic picture. Such novel effects are worth investigating as currently untested predictions of QFT. A similar spirit was recently pushed forward for neutrino oscillations~\cite{Dobrev:2025crc} in the attempt of identifying observable departures from the regular oscillation formula that can be computed in the spacetime collision theory formalism.

\section{Spacetime collision theory}
\label{sec:2}

The physical state of a single scalar particle with non-vanishing mass $m$ is described by a linear combination of plane-wave momentum eigenstates weighted by a wave-packet function $\varphi$. The state vector ket and the corresponding bra are given by
\beq\label{eq:state}
|\varphi\rangle = \frac1{(2\pi)^3}\In{4} \frac{d^3\0{\3{p}}}{2\0{p}^0} \,\varphi(\0{\3{p}})|\0{\3{p}}\rangle\,,\qquad
\langle\varphi| = \frac1{(2\pi)^3}\In{4} \frac{d^3\1{\3{p}}}{2\1{p}^0}\, \varphi^*\hspace{-2pt}(\1{\3{p}})\langle\1{\3{p}}|\,.
\eeq
The equation uses a notation (see also appendix\,\ref{app:notation}) that we employ throughout the paper in order to distinguish the different types of momenta that arise in spacetime collision theory. The momentum of the ket plane wave is marked with an empty circle and the one of the bra plane wave is marked with a black dot. In the calculation of the transition probability, circled and dotted momenta flow in the Feynman amplitude and in the conjugate amplitude, respectively. See later figure\,\ref{diag}.

The four-momenta are represented with regular italic fonts while boldface fonts are used to denote their  3-space components. The two four-momenta associated with the state~(\ref{eq:state}) are on-shell and have positive energy, namely
\beq\label{eq:ltr}
{\0{p}}^\mu=(\0{p}^0=E(\0{\3{p}}),\0{\3{p}})^\mu\,,\qquad {\1{p}}^\mu=(\1{p}^0=E(\1{\3{p}}),\1{\3{p}})^\mu\,,\qquad E(\3{p})=\sqrt{\3{p}^2+m^2}\,.
\eeq
The 3-momentum integrals in eq.\,(\ref{eq:state}) are performed with the standard Lorentz-invariant integration measure. 

We introduce for future convenience the mean and the difference between the circled and dotted momenta
\beq\label{eq:meandifmom}
p\equiv\frac{\0{p}+\1{p}}2\,,\qquad \di{p}\equiv \0{p}-\1{p}\,.
\eeq
The mean momentum is not labelled with any special symbol and the momentum difference is marked with a tilde. Notice that $p^2\geq m^2$ and $p^0\geq m$, while $\di{p}^2\leq0$. Finally, we define the \emph{classical} momentum
\beq\label{eq:cmom1}
P^\mu\equiv m\frac{p^\mu}{\sqrt{p^2}}=(P^0\hspace{-3pt}=\hspace{-2pt}E(\3{P}),\3{P})^\mu,\qquad E(\3{P})=\sqrt{\3{P}^2+m^2}\,,
\eeq
represented with capital fonts.  The $P$ momentum is on-shell and it has positive energy.

The experiment is the repeated observation of states whose wave-packet function fluctuates statistically. In the simplest schematization, the system is described by a discrete (possibly infinite) set of wave-packet functions, $\{\varphi_\alpha(\cdot)\}$, that are realized at each trial with probability $w_\alpha$. We call this system a beam of particles and we define its density matrix
\beq\label{eq:DM}
\rho(\0{\3{p}},\1{\3{p}})=\sum_{\alpha}w_\alpha
\,\varphi^*_\alpha\hspace{-2pt}(\1{\3{p}})
\,\varphi_\alpha\hspace{-1pt}(\0{\3{p}})\,.
\eeq
In a truly physical beam, particles are typically packed in dense bunches. In our schematization instead the particles are sent one-by-one to the collision point. This neglects the possibility of multiple interactions, Beamstrahlung and of any other effect associated with the simultaneous presence of many particles in the collision region.

Before studying collision processes that involve two or several beams it is convenient to consider measurements performed on a single beam.

\subsection{Single-beam measurements
}

The expectation value of a measurement $M$ performed on the beam is the statistical average with weights $w_\alpha$ of the expectation value on the state $|\varphi_\alpha\rangle$ of the corresponding operator:
\beq\label{eq:avg0}
{\mathds{E}}[M]=\sum_\alpha w_\alpha \langle \varphi_\alpha|M|\varphi_\alpha\rangle=
\frac1{(2\pi)^6}\In{4}\In{4}
\frac{d^3\0{\3{p}}}{2\0{p}^0} 
\frac{d^3\1{\3{p}}}{2\1{p}^0} \,
\rho(\0{\3{p}},\1{\3{p}}) \langle\1{\3{p}}|M|\0{\3{p}}\rangle
\,.
\eeq
This displays the  interpretation of $\rho$ as a density matrix: in the calculation of the expectation value it gets multiplied by the matrix element of the quantum operator and traced. Compatibly with the density matrix interpretation, $\rho$ is Hermitian and it has unit trace
\beq
\rho^*\hspace{-2pt}(\1{\3{p}},\0{\3{p}})=\rho(\0{\3{p}},\1{\3{p}})\,,\qquad
\frac1{(2\pi)^3}\In{3}\frac{d^3\0{\3{p}}}{2\0{p}^0}
\,\rho(\0{\3{p}},\0{\3{p}})=1\,.
\eeq
The second equation stems from the normalization of the $w_\alpha$ probabilities and of the wave-packet functions.

If the measurement is a local operator $M=O(x)$ we can use translational symmetry
\beq
\langle\1{\3{p}}|O(x)|\0{\3{p}}\rangle=
{\mathcal{O}}(\0{\3{p}},\1{\3{p}})e^{-i\di{p}\cdot{x}},
\eeq
and write the expectation value as
\beq\label{eq:avgloc}
{\mathds{E}}[{O(x)}]=
\frac1{(2\pi)^6}\In{4}\In{4}
\frac{d^3\0{\3{p}}}{2\0{p}^0} 
\frac{d^3\1{\3{p}}}{2\1{p}^0} \,
\rho(\0{\3{p}},\1{\3{p}}) \,
e^{-i\di{p}\cdot{x}}
{\mathcal{O}}(\0{\3{p}},\1{\3{p}})
\,.
\eeq
A particularly simple local measurement is the one of the charge density operator $Q(x)$, which is the temporal component of the corresponding conserved current. In this case
\beq\label{eq:avgcurr}
{\mathds{E}}[{Q(x)}]=
\frac1{(2\pi)^6}\In{4}\In{4}
\frac{d^3\0{\3{p}}}{2\0{p}^0} 
\frac{d^3\1{\3{p}}}{2\1{p}^0} \,
\rho(\0{\3{p}},\1{\3{p}}) \,
e^{-i\di{p}\cdot{x}}(2p^0)\,
{\mathcal{F}}({\di{p}}^2) 
\,,
\eeq
where ${\mathcal{F}}({\di{p}}^2)$ is the scalar form-factor that parametrises the matrix element of the conserved current as in eq.\,(\ref{eq:me}). Notice that ${\mathcal{F}}({\di{p}}^2)=q$ (with $q$ the charge of the particle) if the particle is free. In a general interacting theory instead
\beq\label{eq:FFq}
{\mathcal{F}}({\di{p}}^2) \simeq q\quad {\rm{if\ }}{\di{p}}^2{\rm{\ is\ small.}}
\eeq
The measurements of other local observables such as the energy and momentum densities and the corresponding fluxes are studied in appendix\,\ref{app:LM}.

\subsection{The Wigner density 
}\label{ss:cl}

The beam is characterised by its density matrix, or in alternative by its \emph{Wigner density}. The latter assumes special significance and is particularly useful to study the classical limit of single-beam measurements and of multi-beam collisions. A classical beam of point-like particles is described by the classical space-momentum density
\beq\label{eq:PCdef}
{\mathcal{P}}_C\hspace{-1pt}(x,\3{P})=\frac{d^6{\rm{Prob}}(t)}{d^3\hspace{-1pt}\3{x}\,d^3\hspace{-2pt}\3{P}}\,,
\eeq
which is the probability to find the particle with 3-momentum $\3{P}$ and space position $\3{x}$, at the time $x^0=t$. The Wigner density, denoted as
\beq
{\mathcal{P}}(x,\3{P})\,,
\eeq
is the quantum analogue of the space-momentum density. We define it as a function of the spacetime position vector $x=(t,\3{x})$ and (unlike previous literature) of the classical 3-momentum introduced in eq.\,(\ref{eq:cmom1}).

The Wigner density is best introduced by comparing the expression for the charge density expectation value~(\ref{eq:avgcurr}) with the classical prediction for the same quantity. In classical physics, the expectation value of a local measurement is the probability that the particle is found at the relevant spacetime position $x$ and with a momentum equal to $\3{P}$, times the value of the corresponding observable when the particle momentum is $\3{P}$, integrated over $\3{P}$. The particle's charge is $q$, independently of $\3{P}$, therefore the classical expectation value of the charge density measurement is simply
\beq\label{eq:avgcurrCL}
{\mathds{E}}_C[{Q(x)}]=q\,
\In{4}d^3\hspace{-2pt}\3{P}\,{\mathcal{P}}_C(x,\3{P}) 
\,.
\eeq
We aim at a definition of the Wigner density by which eq.\,(\ref{eq:avgcurr}) takes the form of a classical expectation value, with ${\mathcal{P}}_C$ replaced by ${\mathcal{P}}$. We will see that this correspondence is possible only provided that the form-factor can be approximated by the charge using eq.\,(\ref{eq:FFq}). The approximation is justified in the classical limit discussed in section\,\ref{ssecCL}.

The classical expectation value is an integral over $\3{P}$, therefore the first step is to change coordinates in the integral of eq.\,(\ref{eq:avgcurr}). The new coordinates are the space components $\3{P}$ and $\di{\3{p}}$ of the classical momentum $P$~(\ref{eq:cmom1}) and of the momentum difference $\di{p}$~(\ref{eq:meandifmom}). We can express the temporal component of $\di{p}$ in terms of the new coordinates $\3{P}$ and $\di{\3{p}}$ by exploiting that $\di{p}$ is orthogonal to $P$, obtaining
\beq\label{eq:dimom1}
\di{p}^0=\frac{\3{P}\cdot\di{\3{p}}}{E(\3{P})}\quad\Rightarrow\quad\0{p}-\1{p}=\di{p}=\left(\frac{\3{P}\cdot\di{\3{p}}}{E(\3{P})},\di{\3{p}}\right),
\eeq
where the dot denotes the Euclidean 3-vectors product. This in turn allows us to compute $p^2$ (the mean momentum squared) in the new coordinates
\beq\label{eq:p21}
p^2=m^2-\di{p}^2/4=m^2+\frac{\di{\3{p}}^2}4-\frac{(\3{P}\cdot\di{\3{p}})^2}{4E^2\hspace{-1pt}(\3{P})}\,.
\eeq
Using this formula we can express $\3{p}$ in  the new coordinates $(\3{P},\di{\3{p}})$ by inverting eq.\,(\ref{eq:cmom1}) and eventually write down (see table\,\ref{tab:tras}) $\0{\3{p}}$ and $\1{\3{p}}$ in the new coordinates.

The integral over $\0{\3{p}}$ and $\1{\3{p}}$ in eq.\,(\ref{eq:avgcurr}) is easily turned into an integral over $\3{P}$ and $\di{\3{p}}$. We first change variables to $\3{p}=(\0{\3{p}}+\1{\3{p}})/2$ and $\di{\3{p}}=\0{\3{p}}-\1{\3{p}}$, which has trivial Jacobian. Next we notice that the residual change of variable from $\3{p}$ to $\3{P}$ is simple because the two vectors are parallel. In spherical coordinates the only non-trivial Jacobian stems from the change from the $|\3{p}|$ to the $|\3{P}|$ variable and these are related by
\beq
|\3{p}|=\frac{\sqrt{p^2}}{m}|\3{P}|\,.
\eeq
 The Lorentz-invariant integration measure of eq.\,(\ref{eq:avgcurr}) is found to be
\beq\label{eq:intmeas}
\frac{d^3\0{\3{p}}}{\0{p}^0}\frac{d^3\1{\3{p}}}{\1{p}^0}=\frac{d^3\hspace{-2pt}\3{P}}{P^0}\,
\frac{\sqrt{p^2}}{m}\frac{d^3\di{\3{p}}}{P^0}
=\frac{d^3\hspace{-2pt}\3{P}}{P^0}\,
\frac{p^2}{m^2 p^0}d^3\di{\3{p}}
.
\eeq
where $P^0=E(\3{P})$ and $p^2$ is given by eq.\,(\ref{eq:p21}). For future convenience, in the second equality we replaced $P^0=m\,p^0/\sqrt{p^2}$.

After the change of variables the expectation value of the charge density becomes
\beq\label{eq:avgcurrCV}
{\mathds{E}}[{Q(x)}]=q\,
\In{4}d^3\hspace{-2pt}\3{P}\left[
\frac1{(2\pi)^6}\frac1{2P^0}
\In{3}d^3\hspace{-1pt}\di{\3{p}}\,
\frac{{\mathcal{F}}({\di{p}}^2)}{q}
\frac{p^2}{m^2}
\rho(\0{\3{p}},\1{\3{p}}) \,
e^{-i\di{p}\cdot{x}}
\right].
\eeq
The only way to reproduce the classical expectation formula~(\ref{eq:avgcurrCL}) would be to define the Wigner density using the expression in the square bracket. However, this is not viable because the form-factor ${\mathcal{F}}$ is specific of the charge density measurement while the Wigner density should offer an observable-independent description of the beam. We thus define the Wigner density as
\beq\label{eq:Wddef}
{\mathcal{P}}(x,\3{P})\equiv
\frac1{(2\pi)^6}\frac1{2P^0}
\In{3}d^3\hspace{-1pt}\di{\3{p}}\,
\frac{p^2}{m^2}
\rho(\0{\3{p}},\1{\3{p}}) \,
e^{-i\di{p}\cdot{x}},
\eeq
obtaining a correspondence between eq.\,(\ref{eq:avgcurrCV}) and the classical formula~(\ref{eq:avgcurrCL}) only under the approximation ${{\mathcal{F}}({\di{p}}^2)}/{q}\simeq1$, using eq.\,(\ref{eq:FFq}). We will see in the next section that the approximation is indeed justified in the classical limit, where ${\di{p}}^2$ can be treated as small. 

The classical space-momentum density is a Lorentz scalar (see e.g.~\cite{degroot1980relativistic}) and the same is true for our definition of the Wigner density. Namely, under a change of frame
\beq\label{eq:LIN}
{\mathcal{P}}( x,\3{P})
\;\longrightarrow\;
{\mathcal{P}}(\Lambda x, \Lambda_m\hspace{-1pt}({\3{P}}))
\,,
\eeq
where $\Lambda$ is a Lorentz transformation, which acts linearly (as $x\to\Lambda x$) on the four-vector $x$. The action of $\Lambda$ on the 3-momentum $\3{P}$---denoted as $\Lambda_m\hspace{-1pt}({\3{P}})$ and reported in eq.\,(\ref{eq:trex})---is non-linear and it corresponds to transforming $\3{P}$ as the 3-space part of a four-momentum vector with positive energy and mass $m$, i.e. with a temporal component $P^0=E(\3{P})$. 

In order to show that eq.\,(\ref{eq:Wddef}) obeys eq.\,(\ref{eq:LIN}) it is sufficient to notice that the $\di{p}$ four-vector is orthogonal to $P$ and we can write
\beq\label{eq:invm}
\frac{d^3\hspace{0pt}\di{\3{p}}}{2P^0}=\delta(2P\cdot\di{p})\,d^4\hspace{-1pt}{\di{p}}\,\;\,\Rightarrow\;\,
{\mathcal{P}}(x,\3{P})=
\frac1{(2\pi)^6}
\In{3}d^4\hspace{-1pt}\di{p}\,\,
\delta(2P\cdot\di{p})
\frac{p^2}{m^2}
\rho(\0{\3{p}},\1{\3{p}}) \,
e^{-i\di{p}\cdot{x}}.
\eeq
The density matrix~(\ref{eq:DM}) is a scalar because the wave-packet functions are scalars and the $w_\alpha$ probabilities are Lorentz-invariant. Therefore the Wigner density is the integral with a Lorentz-invariant measure of a scalar function and hence it is itself a scalar as anticipated. 

Our definition of the Wigner density differs from the one given in ref.~\cite{Kotkin:1992bj} and adopted in the subsequent literature. A comparison is presented in appendix~\ref{sec:relwd}, showing in particular why our definition is a Lorentz scalar while the original definition is not. Other properties of $\mathcal{P}$ that comply with the expected properties of a classical space-momentum density are
\beq\label{eq:WDP}
{\mathcal{P}}^*\hspace{-2pt}( x,\3{P})=
{\mathcal{P}}( x,\3{P}),\qquad
\In{4}\In{4}
d^3\hspace{-1pt}{\3{x}} \, d^3\hspace{-2pt}{\3{P}}\,{\mathcal{P}}( x,\3{P})=1\,,
\qquad P^\mu\partial_\mu {\mathcal{P}}( x,\3{P})=0\,.
\eeq
The first two properties follow from the Hermiticity and normalization of $\rho$, respectively. The last equation stems from the fact that $P^\mu$ is orthogonal to $\di{p}^\mu$ and it matches the conservation equation that drives the time evolution of the space-momentum density of a free classical particle. Notice however that  ${\mathcal{P}}$, in spite of the many analogies, is not a classical probability density because it is not positive in general.

\subsection{The classical beam limit}\label{ssecCL}

The beams employed for real collider experiments are accurately described by classical physics and quantum effects are present only in the short-distance interactions among the particles. This indicates heuristically that real beams are in a regime where the underlying quantum reality is effectively classical to an excellent approximation. In this section we identify the condition under which the single-beam measurements predictions approach the classical result, providing a precise definition of the \emph{classical limit} in our context. In section\,\ref{se:sds} we show that the standard approach to collision theory also emerges in the same limit as an accurate description of the collision among stable particles with short-distance interactions.

Eq.\,(\ref{eq:avgloc}) is the expectation value of the measurement of a local observable. Its quantum nature stems from the fact that the observable ${\mathcal{O}}(\0{\3{p}},\1{\3{p}})$ depends on two momenta $\0{\3{p}}$ and $\1{\3{p}}$ while a classical point-like particle is characterised by only one momentum. In a classical calculation, the expectation value of a local measurement performed at a spacetime point $x$ receives contributions only from those particles whose trajectory intercepts $x$. As the position is fixed, the state of the particle and in turn the value of the relevant observable is fully specified by its 3-momentum. Therefore, the classical result for the expectation value of a local measurement is an observable that depends on a single momentum, averaged over the observable-independent particle distribution. This structure emerges in eq.\,(\ref{eq:avgloc}) only if the density matrix $\rho(\0{\3{p}},\1{\3{p}})$ is strongly peaked around $\0{\3{p}}=\1{\3{p}}$, in such a way that $\0{\3{p}}\simeq\1{\3{p}}$ on the domain of integration and ${\mathcal{O}}(\0{\3{p}},\1{\3{p}})$ effectively depends on one momentum only. Unsurprisingly, the classical limit is when the density matrix is nearly diagonal.

For a Lorentz-invariant characterization of a classical beam we can employ $-\di{p}^2$ (which is positive-definite), where $\di{p}$ is the momentum difference four-vector defined in eq.\,(\ref{eq:meandifmom}). The support of the density matrix of a nearly classical beam is in a region of the $(\0{\3{p}},\1{\3{p}})$ space where $-\di{p}^2$ is small:
\beq\label{eq:supp}
{\rm{supp}}\big[\rho(\0{\3{p}},\1{\3{p}})\big]\subset\big\{-\di{p}^2\lesssim\kappa^2\big\}\,.
\eeq
This equivalently means that $\rho$ is zero or exponentially suppressed if $-\di{p}^2$ is larger or much larger than a small parameter $\kappa^2$ (see later) that characterizes the beam. We notice that 
\beq
-\di{p}^2=\di{\3{p}}^2-\frac{(\di{\3{p}}\cdot\3{P})^2}{E^2\hspace{-1pt}(\3{P})}\geq \di{\3{p}}^2\left(
1-\frac{\3{P}^2}{E^2\hspace{-1pt}(\3{P})}
\right)
=\di{\3{p}}^2
\frac{m^2}{E^2\hspace{-1pt}(\3{P})}
\,,
\eeq
where the lower bound is attained when $\di{\3{p}}$ is parallel to $\3{P}$. By this relation we can bound the size of the individual components of $\di{p}$ in any reference frame
\al{\label{eq:momtilim}
&|\di{\3{p}}|\leq E\hspace{-1pt}(\3{P})\frac{\sqrt{-\di{p}^2}}{m}\leq E\hspace{-1pt}(\3{P})\frac{\kappa}{m}\,,
\nonumber\\
&|\di{p}^0|\leq
|\di{\3{p}}|\frac{|\3{P}|}{E\hspace{-1pt}(\3{P})}
\leq |\3{P}| \frac{\sqrt{-\di{p}^2}}{m}\leq 
|\3{P}| \frac{\kappa}{m}\,.
}

This shows that in the classical limit ($\kappa\to0$) $\di{p}=\0{p}-\1{p}$ vanishes in any frame such that the circled and dotted momenta coincide and become equal to the mean momentum. The mean momentum approaches the classical momentum up to quadratic corrections because
\beq\label{eq:momcllim}
{\rm{abs}}\hspace{-2pt}\left({p}^\mu-P^\mu\right)
=\big(\sqrt{p^2}/m-1\big)\,{\rm{abs}}\hspace{-2pt}\left(P^\mu\right)
\leq \frac{-\di{p}^2}{8m^2} {\rm{abs}}\hspace{-2pt}\left(P^\mu\right) \lesssim\frac{\kappa^2}{8m^2} {\rm{abs}}\hspace{-2pt}\left(P^\mu\right),
\eeq
where we employed eq.\,(\ref{eq:p21}).

The formal classical limit is $\kappa\to0$, however $\kappa$ is finite for a real beam entailing corrections to the classical approximations that need to be estimated. In a theory without any massless particle we can define the mass gap $\mu\neq0$ as the mass of the lightest particle. The classical approximation holds provided
\beq\label{eq:dmsmu}
\kappa\ll\mu\,,
\eeq
and the corrections to single-beam measurements are polynomial in $\kappa/\mu$. This is easy to establish in the case of the charge density expectation value calculation illustrated in the previous section. The electromagnetic form-factor is analytic and it can be expanded as
\beq\label{eq:ffe}
{\mathcal{F}}(\di{p}^2)={\mathcal{F}}(0)+\frac{\di{p}^2}{\mu^2}{\mathcal{F}}^\prime(0)+\ldots\,.
\eeq
The first term is equal to the charge~(\ref{eq:FFq}) and it produces a classical result of the form of eq.\,(\ref{eq:avgcurrCL}). The other terms give corrections suppressed by $-\di{p}^2/\mu^2\lesssim\kappa^2/\mu^2$ to the expectation value~(\ref{eq:avgcurrCV}). In appendix~\ref{app:LM} we compute the expectation value of a variety of local measurements and we check that the classical result is indeed recovered up to quadratic corrections in all cases. The corrections are genuinely non-classical contributions to the expectation value. It is easy to see from the definition~(\ref{eq:Wddef}) that these contributions are proportional to spacetime derivatives (specifically, Lorentz-invariant boxes) of the Wigner density and for this reason they do not possess a classical probabilistic interpretation.

Real collider experiments beams feature a very small $\kappa$ and this fact justifies the validity of the classical approximation. Indeed, $\di{p}$ is in essence---see the definition of the Wigner density in eq.\,(\ref{eq:Wddef})---the Fourier conjugate variable of the position $x$ and $x$ is delocalized over macroscopic length scales. For instance, the LHC beam size at the interaction point is of order $10\,\mu{\rm{m}}$ in the transverse direction and much larger in the longitudinal direction. The localization of its Wigner density~(\ref{eq:Wddef}) in space is thus $\Delta{\3{x}}\sim 10\,\mu{\rm{m}}$, which means in Fourier space $\kappa\sim1/\Delta{\3{x}}=0.02\,{\rm{eV}}$. Any process with an effective mass-gap $\mu$ that is larger than this tiny threshold is accurately described by the classical limit. This estimate assumes that the Wigner density is a smooth function, otherwise the inverse of the beam size is not an accurate estimate of the extension of the density matrix in the $\di{p}$ space.

The situation is very different if a massless particle such as the photon is present in the theory and its effects cannot be ignored in the calculation because in this case there is not a physical $\mu$ scale that controls the $\di{p}^2$ corrections to the form-factor. In QED, the derivatives of the form-factors at $\di{p}^2=0$  feature IR loop divergences that are not cancelled by any real contribution. Our treatment fails because the physical system is not described by single-particle states like in eq.\,(\ref{eq:state}), but the particle is instead accompanied by infinitely many soft photon quanta~\cite{Kulish:1970ut}. The spacetime collision formalism was originally introduced~\cite{Kotkin:1992bj} to model finite beam size QED effects at $e^+e^-$ colliders, but it is not a rigorous approach to this problem because it is based on eq.\,(\ref{eq:state}). The caveat does not apply when the formalism is used for processes with a mass-gap that are the focus of the present article.

\subsection{Collision processes}

Collision processes occur in the presence of two or more single-particle beams. The calculation of the transition probability is in principle straightforward, starting from the elementary formula
\beq\label{eq:Tm}
\langle out | iT | in \rangle=(2\pi)^4\delta^4(p_{out}-p_{in})\,i{\mathcal{M}}\,,
\eeq
for the elements of $T$-matrix ($S=\mathds{1}+iT$) between plane-wave states. In the equation, $in$ and $out$ are respectively the initial and the final state, with total 4-momenta $p_{in}$ and $p_{out}$. The Feynman amplitude for the $in\to out$ process is denoted as $i{\mathcal{M}}$. Both connected and disconnected Feynman diagrams contribute to the amplitude, a priori.

The $n$ colliding particles in the initial state are described by wave-packets $\varphi_1,\ldots,\varphi_n$ that fluctuate independently at each repetition of the collision experiment. The physical nature of the final state particles needs not to be specified as long as we are interested in computing only the {total probability} for the production of a certain set of final state particles and not their kinematic distributions. In this case, an arbitrary basis (in particular, plane waves) can be employed to describe the final state. The transition probability reads\,\footnote{This treatment is valid also for differential rates in sufficiently large bins and in general for observables that are insensitive to the coherence between different final state particle momenta.} 
\beq\label{eq:probability}
p=\sum_{\alpha_1\ldots \alpha_n}\hspace{-3pt}w_{\alpha_1}\ldots w_{\alpha_{n}} \prod\limits_{f=1}^{n_F}\,\,\In{5}\frac{d^3\hspace{-1pt}\3{k}_f}{(2\pi)^32k^0_f}
\left|
\langle
\3{k}_1\ldots \3{k}_{n_F}|iT|
\varphi_{1,\alpha_1}\ldots \varphi_{n,\alpha_n}\rangle\right|^2,
\eeq
where $k_f$ are the on-shell momenta of the particles in the final state.

We now insert the plane-wave decomposition~(\ref{eq:state}) of the $n$ incoming states. The wave-packet factors and the statistical average combine---see eq.\,(\ref{eq:DM})---into the density matrices $\rho_1,\ldots,\rho_n$ that characterise each colliding beam. We are left with $n$ six-dimensional integrals over the initial state momenta, with volume element
\beq\label{eq:inint}
\d^6\hspace{-2pt}I_i\equiv
\frac{d^3\0{\3{p}}_i}{2\0{p}^0_i} 
\frac{d^3\1{\3{p}}_i}{2\1{p}^0_i} \,
\frac{\rho_i(\0{\3{p}}_i,\1{\3{p}}_i)}{(2\pi)^6}\,.
\eeq
Using eq.\,(\ref{eq:Tm}) we evaluate the $T$-matrix elements obtaining
\al{
p=
\In{4}
\d^6\hspace{-2pt}I_1\ldots
\In{4}
\d^6\hspace{-2pt}I_n
\prod\limits_{f=1}^{n_F}\,\,\In{5}\frac{d^3\hspace{-1pt}\3{k}_f}{(2\pi)^32k^0_f}
&
(2\pi)^4\delta^4\hspace{-2pt}\bigg(k_{\rm{tot}}-\sum_{i=1}^n\0{p}_i\bigg)
(2\pi)^4\delta^4\hspace{-2pt}\bigg(k_{\rm{tot}}-\sum_{i=1}^n\1{p}_i\bigg)
\nonumber\\
\times&{\mathcal{M}}\big(\{\0{p}\};\hspace{-1pt}\{k\}\big)\,
{\mathcal{M}}^*\hspace{-2pt}\big(\{\1{p}\};\hspace{-1pt}\{k\}\big).\label{eq:Mas1}
}
where $k_{\rm{tot}}$ is the sum of the final particles four-momenta $k_f$.

The left panel of figure\,\ref{diag} offers a pictorial illustration of eq.\,(\ref{eq:Mas1}). The initial particles are characterized by two incoming momenta $\0{p}_i$ and $\1{p}_i$, while a single outgoing momentum $k_f$ is enough to describe each final state particle. The circled $\0{p}_i$ momenta flow into the on-shell Feynman amplitude ${\mathcal{M}}\big(\{\0{{p}}\};\hspace{-1pt}\{{k}\}\big)$, which therefore depends on the set of all the circled momenta---denoted as $\{\0{{p}}\}$---and on the final momenta $\{{k}\}$. The dotted momenta flow into the conjugate amplitude ${\mathcal{M}}^*\hspace{-2pt}\big(\{\1{{p}}\};\hspace{-1pt}\{{k}\}\big)$. We employ a notation where a semicolon divides the incoming from the outgoing amplitude momenta. There are two delta-functions of total momentum conservation in eq.\,(\ref{eq:Mas1}), which set both the sum of the circled momentum and the sum of the dotted momenta to be equal to the total momentum of the final state.

\input{Figures/fig1.tex}

A more compact expression for the transition probability is obtained using the mean momenta and the momentum difference of the incoming particles. Their definition~(\ref{eq:meandifmom}) is reported on the right panel of figure\,\ref{diag}. Momentum conservation holds separately for the circled and dotted momenta, with the same final state total momentum $k_{\rm{tot}}$. Hence the sum of the mean momenta $p_{\rm{tot}}=\sum_ip_i$ equals $k_{\rm{tot}}$ and the sum of the momentum differences $\di{p}_i$ vanishes as indicated in the figure. The delta-functions in eq.\,(\ref{eq:Mas1}) can be turned into the delta-functions that enforce these two conditions. The one enforcing the conservation of the $p$ momenta naturally combines with the volume elements of the $\3{k}_f$ integrals and it reconstructs the phase-space volume element
\beq
\d{\rm{Ph}}(p_{\rm{tot}})=\prod\limits_{f=1}^{n_F}\frac{d^3\hspace{-1pt}\3{k}_f}{(2\pi)^32k^0_f}
(2\pi)^4\delta^4\hspace{-1pt}(k_{\rm{tot}}-p_{\rm{tot}})\,.
\eeq
If we now define the \emph{collision function}
\beq\label{eq:sF}
{\mathfrak{C}}(\{{\0{{p}}}\},\{{\1{{p}}}\})\equiv\In{4}\d{\rm{Ph}}(p_{\rm{tot}})\,
{\mathcal{M}}\big(\{\0{{p}}\};\hspace{-1pt}\{{k}\}\big)\,
{\mathcal{M}}^*\hspace{-2pt}\big(\{\1{{p}}\};\hspace{-1pt}\{{k}\}\big),
\eeq
we obtain a compact expression for the transition probability
\beq\label{eq:MAS}
p=
\In{4}
\d^6\hspace{-2pt}I_1\ldots
\In{4}
\d^6\hspace{-2pt}I_n
(2\pi)^4\delta^4\hspace{-2pt}\bigg(\sum_{i=1}^n\di{p}_i\bigg)\,
{\mathfrak{C}}(\{{\0{{p}}}\},\{{\1{{p}}}\}).
\eeq

The collision function $\mathfrak{C}$ carries all the dependence on the microscopic process under examination including the final state particles phase-space and the Feynman amplitude. The transition probability is given by the function $\mathfrak{C}$ integrated over the initial state beams with integration volume element~(\ref{eq:inint}) that depends on their density matrices $\rho_i$. The integral is constrained by the delta-function enforcing the conservation (specifically, that they sum up to zero) of the momentum differences $\di{p}_i$. 

We can rewrite eq.\,(\ref{eq:MAS}) in a more explicit form that is particularly suited for studying its classical limit. By eq.\,(\ref{eq:intmeas}) we change variables in the integration elements, obtaining
\beq
\d^6\hspace{-2pt}I_i=
{d^3\hspace{-2pt}\3{P}_i}
\,
\frac1{(2\pi)^6}\frac1{2P^0_i}
{d^3\hspace{-1pt}\di{\3{p}}_i}
\frac{p^2_i}{m_i^2}\rho_i(\0{\3{p}}_i,\1{\3{p}}_i)\frac1{2p^0_i},
\eeq
where $P_i$ are the classical momenta of the colliding beams defined as in eq.\,(\ref{eq:cmom1}). With the new variables eq.\,(\ref{eq:MAS}) becomes
\beq\label{eq:MAS01}
p=
\prod\limits_{i=1}^{n}\hspace{-2pt}\bigg(
\In{3}{d^3\hspace{-2pt}\3{P}_i}
\frac1{(2\pi)^6}\frac1{2P^0_i}
\In{3}
{d^3\hspace{-1pt}\di{\3{p}}_i}
\frac{p^2_i}{m_i^2}\rho_i(\0{\3{p}}_i,\1{\3{p}}_i)
\bigg)
(2\pi)^4\delta^4\hspace{-2pt}\bigg(\sum_{i=1}^n\di{p}_i\bigg)\,\frac{{\mathfrak{C}}(\{{\0{{p}}}\},\{{\1{{p}}}\})}{
\prod_i(2p_i^0)
}.
\eeq

A last useful step is to express the delta-function using the Fourier representation, which produces an integral over a spacetime position four-vector $x$ and a product of oscillatory phases for each beam. We obtain
\beq\label{eq:MAS1}
p=\In{3}d^4\hspace{-1pt}x
\prod\limits_{i=1}^{n}\hspace{-2pt}\bigg(
\In{3}{d^3\hspace{-2pt}\3{P}_i}
\left[
\frac1{(2\pi)^6}\frac1{2P^0_i}
\In{3}
{d^3\hspace{-1pt}\di{\3{p}}_i}
\frac{p^2_i}{m_i^2}\rho_i(\0{\3{p}}_i,\1{\3{p}}_i)
e^{-i\di{p}_i\cdot{x}}
\right] \hspace{-2pt}\bigg)
\frac{{\mathfrak{C}}(\{{\0{{p}}}\},\{{\1{{p}}}\})}{
\prod_i(2p_i^0)
},
\eeq
where we collected into square brackets the factors that match the definition of the Wigner density given in eq.\,(\ref{eq:Wddef}). Notice however that the corresponding integrals do not factorise and they cannot yet be performed because the collision function ${\mathfrak{C}}$ depends on the momentum-difference variables $\di{\3{p}}_i$.

\subsection{Two-particle short-distance collision}
\label{se:sds}

We now specialize to collisions between two particles, $\u$ and $\s$, producing an arbitrary final state. The naming anticipates the next section where $\u$ is regarded as an unstable particle while $\s$ can be stable. 

\subsubsection*{Kinematics}

The initial state is described by a total of twelve variables that can be identified with the independent components of the on-shell circled and dotted momenta $\0{p}_{\u,\s}$ and $\1{p}_{\u,\s}$ or with those of the mean momenta $p_{\u,\s}$ and momentum difference $\di{p}_{\u,\s}$, or finally with the classical 3-momenta $\3{P}_{\u,\s}$ and 3-momentum differences $\di{\3{p}}_{\u,\s}$. However, momentum conservation produces a $\delta^4(\di{p}_{\u}+\di{p}_{\s})$ in the transition probability~(\ref{eq:MAS01}) that restricts the momenta by the four conditions
\beq\label{eq:ptcons}
\di{p}_{\u}=-\di{p}_{\s}\equiv\di{p}\,.
\eeq

Only eight variables are thus sufficient to characterize the kinematic of the two incoming beams. Using eq.\,(\ref{eq:dimom1}), the tilded momentum conservation equation~(\ref{eq:ptcons}) gives
\beq
\di{\3{p}}_{\u}=-\di{\3{p}}_{\s}=\di{\3{p}}\,,\;\qquad\;
\Delta\vec{v}\cdot\di{\3{p}}=0\,,
\eeq
where $\Delta\vec{v}$ is the difference between the classical velocities of the incoming particles
\beq\Delta\vec{v}=
\frac{\3{P}_\u}{E(\3{P}_\u)}-\frac{\3{P}_\s}{E(\3{P}_\s)}.
\eeq
The eight independent variables can be chosen as the classical 3-momenta $\3{P}_{\u,\s}$ of the two particles plus the 3-momentum difference variable $\di{\3{p}}$ that however spans a two-dimensional space as it is constrained to be orthogonal to the velocity difference. For future convenience we report here the $\di{\3{p}}$ integration measure of eq.\,(\ref{eq:MAS01}) after the change of variables
\beq\label{eq:vintpt}
\delta^4\hspace{-1pt}(\di{p}_\u+\di{p}_\s)\,d^3\hspace{-1pt}\di{\3{p}}_\u d^3\hspace{-1pt}\di{\3{p}}_\s =\delta(\Delta\vec{v}\cdot\di{\3{p}})\,d^3\hspace{-1pt}\di{\3{p}}=
\frac{d^2\hspace{-1pt}\di{\3{p}}}{|\Delta\vec{v}|}\,.
\eeq

The collision function is Lorentz invariant, hence it depends only on the two  invariants that remain independent after imposing the constraint~(\ref{eq:ptcons}). These are 
\beq\label{eq:meanS}
s\equiv (p_\u+p_\s)^2=p_{\rm{tot}}^2\,,\qquad\di{p}^2=\di{p}^2_\u=\di{p}^2_\s\,.
\eeq
Possible invariants constructed as products of $p$'s with $\di{p}$'s vanish because $p$ and $\di{p}$ of the same particle are orthogonal and $\di{p}_\u$ is equal to $-\di{p}_\s$. The other invariants $p_{\u,\s}^2$ are functions of $\di{p}^2$ (and the masses) because of eq.\,(\ref{eq:p21}). 

The centre-of-mass energy squared of the mean momenta, $s$, can be traded for the one of the classical momenta, $S\equiv(P_\u+P_\s)^2$. The latter variable is more convenient than $s$ because it stays constant in the classical limit. The two variables are related by
\beq\label{eq:ClS}
\frac{S-m_\u^2-m_\s^2}{m_\u m_\s}=
\frac{s-m_\u^2-m_\s^2+\di{p}^2/2}{\sqrt{p_\u^2p_\s^2}}\,,\;\;{\rm{with}}\;\;\;p_{\u,\s}^2=m_{\u,\s}^2-\di{p}^2/4\,.
\eeq
Notice that $s$ is bounded by the square of the sum of the effective masses of the $p_\u$ and $p_\s$ momenta, which depend on $\di{p}$. The bound on $S$ is instead the squared sum of the $\u$ and $\s$ physical masses. The two constraints are equivalent
\beq\label{eq:bons}
s\geq\bigg(\sqrt{p_\u^2}+\sqrt{p_\s^2}\bigg)^2\quad\Leftrightarrow\quad S\geq(m_\u+m_\s)^2.
\eeq

By expressing ${{\mathfrak{C}}}$ as a function of $S$ and $-\di{p}^2$ (we prefer $-\di{p}^2$ because it is positive) and specializing to $n=2$ we obtain from eq.\,(\ref{eq:MAS1}) the master formula for two-particle collisions
\al{
\label{eq:MAS3}
p=&\In{3}d^4\hspace{-1pt}x
\In{3}{d^3\hspace{-2pt}\3{P}_\u}
\In{3}
{d^3\hspace{-2pt}\3{P}_\s}\Bigg\{\nonumber\\
&\left[
\frac1{(2\pi)^6}\frac1{2P^0_\u}
\In{3}
{d^3\hspace{-1pt}\di{\3{p}}_\u}
\frac{p^2_\u}{m_\u^2}\rho_\u(\0{\3{p}}_\u,\1{\3{p}}_\u)
e^{-i\di{p}_\u\cdot{x}}
\right] \left[
\frac1{(2\pi)^6}\frac1{2P^0_\s}
\In{3}
{d^3\hspace{-1pt}\di{\3{p}}_\s}
\frac{p^2_\s}{m_\s^2}\rho_\s(\0{\3{p}}_\s,\1{\3{p}}_\s)
e^{-i\di{p}_\s\cdot{x}}
\right] \nonumber \\
&\hspace{90pt}\times
\frac{{\mathfrak{C}}(S,-\di{p}^2)}{
4p_\u^0p_\s^0
}\,\Bigg\}.
}
The terms in the square brackets reconstruct the Wigner densities for two beams, but the corresponding integrals cannot yet be performed because ${\mathfrak{C}}$ depends on the integration variables $\di{\3{p}}_\u$ and $\di{\3{p}}_\s$. This dependence emerges exclusively from the dependence on $-\di{p}^2$ because $S$ is only a function of the classical momenta $\3{P}_\u$ and $\3{P}_\s$.

\subsubsection*{Short-distance collision}

The standard description of collisions treats the beams as classical point-like particles that propagate freely until when their trajectories cross and the particles undergo a \emph{short-distance} (exactly localized in space and instantaneous in time) scattering reaction. As this description is extremely successful in practice, it must emerge from the complete collision theory as an accurate description of real experiments. We argued in section\,\ref{ssecCL} that the beams of real collider experiments are effectively classical. We now study how and under what conditions the standard formula for the transition probability emerges taking the classical limit of eq.\,(\ref{eq:MAS3}). 

If the colliding beams are nearly classical, their density matrices are localized in a region where $-\di{p}_{\u}^2$ and $-\di{p}_{\s}^2$ are bounded by the small parameter $\kappa^k$ as in eq.\,(\ref{eq:supp}). Hence the classical approximation corresponds to the expansion of the collision function ${\mathfrak{C}}$ around $-\di{p}^2=0$ at fixed $S$. We now show that the standard short-distance description of the collision is recovered if ${\mathfrak{C}}$ at fixed $S$ is analytic at $\di{p}=0$ in such a way that it can be expanded as 
\beq\label{eq:Sapp}
{\mathfrak{C}}(S,-\di{p}^2)={\mathfrak{C}}(S,0)+\sum\limits_{n>0} c_n(S)\frac{(-\di{p}^2)^n}{\mu^{2n}}\,,
\eeq
where $\mu$ is a finite scale that controls the convergence radius of the expansion.

When $\di{p}=0$, the tilded momenta vanish and the circled, dotted and mean momenta of the particles become equal to the corresponding classical momenta. Therefore the collision function~(\ref{eq:sF}) becomes proportional to the standard scattering cross section $\sigma$
\beq\label{eq:scxs}
{\mathfrak{C}}(S,0)=
\In{4}\d{\rm{Ph}}(P_{\rm{tot}})
\big|{\mathcal{M}}\big(P_\u,P_\s;\hspace{-1pt}\{k\}\big)\big|^2
={\scriptstyle{4\sqrt{(P_\u\cdot P_\s)^2-m_\u^2m_\s^2}}}\;\sigma(S).
\eeq
This zero-order term of the ${\mathfrak{C}}$ expansion is independent of $\di{p}$ and it can be brought outside the $d^3\hspace{-1pt}\di{\3{p}}_{\u,\s}$ integrals in the transition probability\,(\ref{eq:MAS3}). The integrals can be performed and they reconstruct the Wigner densities of the two beams. At order zero in the classical expansion we thus recover the textbook formula\,\footnote{For instance, eq.\,(12.7) of~\cite{Landau:1975pou} shows the probability of a collision happening inside a 4-volume $d^4\hspace{-2pt}x\hspace{-1pt}=\hspace{-1pt}dtdV$, for monochromatic beams. To obtain eq.\,(\ref{eq:MASCL2p}) identify $n_{1,2}$ with ${\mathcal{P}}_{\u,\s}d^3\hspace{-2pt}\3{P}_{\u,\s}$ and integrate. 
}
\al{\label{eq:MASCL2p}
p_0=\hspace{-1pt}\In{3}d^4\hspace{-1pt}x
\In{4}{d^3\hspace{-2pt}\3{P}_\u}\hspace{-2pt}\In{4}{d^3\hspace{-2pt}\3{P}_\s}
{\mathcal{P}}_\u\hspace{-1pt}(x,\3{P}_\u)
{\mathcal{P}}_\s\hspace{-1pt}(x,\3{P}_\s)
\,{\tilde{v}}_{\rm{rel}}\,\sigma(S)\,,&\\
\quad {\tilde{v}}_{\rm{rel}}\equiv
\frac{\scriptstyle\sqrt{(P_\u\cdot P_\s)^2-m_\u^2m_\s^2}}{\scriptstyle P^0_\u P^0_\s}\,.&
\nonumber}

Eq.\,(\ref{eq:MASCL2p}) matches the standard  description of the collision process as due to short-distance scattering. The Wigner density is classically interpreted as the probability density in $\3{x}$ and $\3{P}$ and its evolution in time---see eq.\,(\ref{eq:WDP})---is the one of a beam of free classical particles. This corresponds to a description where the two particles do not interact before the short-distance scattering reaction. The two densities are evaluated at the same space-time point $x$ because the reaction occurs (at the point $x$) only when the particles overlap. The probability that one pair of particles scatters is encapsulated in the cross section (times ${\tilde{v}}_{\rm{rel}}$), which depends on the particles' momenta through $S$. The transition probability is thus averaged over the momenta and finally integrated over the point $x$ where the collision can take place.

Higher-order terms in the ${\mathfrak{C}}$ expansion~(\ref{eq:Sapp}) give corrections to the transition probability that cannot be cast in the form of eq.\,(\ref{eq:MASCL2p}) and in particular they are not described by a cross section. Still they feature a short-distance structure because they are polynomials in $-\di{p}^2$ and hence---notice that the corrections to the $4p_\u^0p_\s^0$ denominator are also polynomials in $-\di{p}^2$~(\ref{eq:momcllim})---they give $(-\di{p}^2)^n$ contributions to the integrand of eq.\,(\ref{eq:MAS3}). Recalling that $-\di{p}^2\hspace{-2pt}=\hspace{-2pt}-\di{p}_\u^2\hspace{-2pt}=\hspace{-2pt}-\di{p}_\s^2$, these terms can be written as $\Box^n$ acting on any of the two exponentials. After the $d^3\hspace{-1pt}\di{\3{p}}_{\u,\s}$ integrals they give $\Box^n$ acting on one of the Wigner densities. Namely, their contribution to the transition probability is not proportional to the product of the Wigner densities but rather to
\beq
{\mathcal{P}}_\s(x,\3{P}_\s)\times
\Box^n{\mathcal{P}}_\s(x,\3{P}_\u)\,.
\eeq
Given that $\Box$ is a local operation, these terms are short-distance effects. They do not possess a classical probabilistic interpretation because they involve derivatives of the probability density.

In theories with a mass-gap $\mu$ it is easy to establish the validity of eq.\,(\ref{eq:Sapp}) for a large class of processes. This equation is the direct analogue of the form-factor expansion~(\ref{eq:ffe}). A case where eq.\,(\ref{eq:Sapp}) is violated, central to the present work, is when at least one of the colliding particles (say, $\u$) is unstable and the Feynman amplitude contains a diagram that combines the diagram of an allowed decay process for the $\u$ particle with the one of an allowed scattering of its decay products. We will see in the next section that this produces non-analyticities in $\mathfrak{C}$ which in turn produce \emph{long distance} terms in the transition probability.  

\newpage
\section{Unstable particle collisions}
\label{sec:3}

We turn now to the study of a general class of collision processes initiated by an unstable particle $\u$, which can be constructed as follows. Consider any of the $\u$ allowed decays and interpret it as the production of one particle $\i$ (arbitrarily selected among the decay products) plus a single- or multi-particle state $\D$ that collects the other decay products
\beq\label{eq:decc}
\u\longrightarrow\i\,\D\,.
\eeq
Next, consider any particle $\s$ that can scatter with $\i$ producing some final state $\P$
\beq\label{eq:prod}
\i\,\s\longrightarrow\P\,.
\eeq
Combining these elements we obtain a process
\beq
\u\,\s\longrightarrow\D\;\P\,,
\eeq
whose Feynman amplitude receives contributions from diagrams with the topology represented in figure\,\ref{diag1}, where the $\i$ particle propagator in the $t$-channel connects the amplitudes that are responsible for the decay (${\mathcal{M}}_\D$) and scattering (${\mathcal{M}}_\P$) processes. 

\input{Figures/fig2.tex}

Diagrams with the topology of figure\,\ref{diag1} are interesting because they feature a divergent $\i$-particle propagator when the momentum flowing in the $t$-channel reaches the mass shell of $\i$. If this is allowed to happen on the physical phase-space, it produces an IR divergence. The kinematic conditions on the masses and scattering energy for the $t$-channel momentum to go on-shell with positive energy (if it is oriented from $\D$ to $\P$ like in the figure) are the exact same conditions that allow simultaneously for the decay~(\ref{eq:decc}) and the scattering~(\ref{eq:prod}) processes. This suggests that the resolution of the divergence is related to a factorized contribution to the collision process where the $\u$ decay is followed by the scattering process initiated by its decay product $\i$ interacting with $\s$. A similar picture emerges when the on-shell $t$-channel momentum has negative energy, which corresponds to a physical $\i$ particle propagating from $\P$ to $\D$. In this case, the interpretation is that the $\i$ particle comes from the $\s$ decay and it scatters on $\u$.

In sections\,\ref{sec:sdt},~\ref{sec:ldt} and~\ref{sec:exc} we show how this factorized interpretation emerges in the spacetime collision theory framework. The finite size of the beam regulates the divergence enabling the determination of the transition probability. The regulated divergence induces terms in the collision function that are structurally different from those we encountered in section\,\ref{se:sds}. The latter terms are analytic and they produce short-distance contributions to the transition probability. The new ones are non-analytic and produce long-distance effects in the transition probability. Specifically, they describe the $\i$ particle produced in the decay travelling a macroscopic distance before scattering. The different spacetime structure of the non-analytic (long-distance) and analytic (short-distance) terms can be visualized by comparing the middle with the right panel of figure\,\ref{fig:one}.

\subsection{The collision function}
\label{sec:sdt}

In section\,\ref{se:sds} we discussed general properties of the $\u$-$\s$ collision process including the parametrization of the kinematics in terms of a single tilded momentum $\di{p}\hspace{-1pt}=\hspace{-1pt}\di{p}_\u\hspace{-2pt}=\hspace{-2pt}-\di{p}_\s$ and a study of the classical limit under the assumption that the collision function is analytic at $\di{p}=0$. In the present and the following section we isolate the non-analyticities due to the diagrams with the topology of figure\,\ref{diag1} and study their effect on the transition probability in the classical limit. The starting point is the definition of the collision function~(\ref{eq:sF}) specialized to collisions with two particles in the initial state
\beq\label{sf2ps}
{\mathfrak{C}}(S,-\di{p}^2)\hspace{-1pt}=\hspace{-1pt}\In{4}\d{\rm{Ph}}(p_{\rm{tot}})\,˜
{\mathcal{M}}\big({{p}}_\u\hspace{-2pt}+\hspace{-1pt}\di{{p}}/2,{p}_\s\hspace{-2pt}-\hspace{-1pt}\di{{p}}/2;\{{k}\}\hspace{-1pt}\big)\,
{\mathcal{M}}^*\hspace{-2pt}\big({p}_\u\hspace{-2pt}-\hspace{-1pt}\di{{p}}/2,{p}_\s\hspace{-2pt}+\hspace{-1pt}\di{{p}}/2;\{{k}\}\hspace{-1pt}\big),
\eeq
where $\{{k}\}$ denotes collectively the final state particles momenta over which the phase-space integral is performed. The total momentum $p_{\rm{tot}}$ on the phase-space can be expressed as $p_{\rm{tot}}\hspace{-2pt}=\hspace{-2pt}p_\u\hspace{-1pt}+\hspace{-1pt}p_\s$ or equivalently $p_{\rm{tot}}\hspace{-2pt}=\hspace{-1pt}\0{p}_\u\hspace{-1pt}+\hspace{-1pt}\0{p}_\s\hspace{-1pt}=\hspace{-1pt}\1{p}_\u\hspace{-1pt}+\hspace{-1pt}\1{p}_\s$ because the sum of the tilded momenta vanishes. 

The $\i$ particle propagator of figure\,\ref{diag1} contributes to the integrand with a factor
\beq\label{eq:Pidef}
\Pi\equiv\frac1{\0{q}^2-m_\i^2+i\epsilon}\frac1{\1{q}^2-m_\i^2-i\epsilon},
\eeq
where $\0{q}^2$ and $\1{q}^2$ are the $t$-channel variables of the circled and dotted kinematics, namely
\al{
\0{q}^2=m_\u^2+\mu_\D^2-2\0{p}_\u\cdot K_\D=m_\s^2+\mu_\P^2-2\0{p}_\s\cdot K_\P,
\\
\1{q}^2=m_\u^2+\mu_\D^2-2\1{p}_\u\cdot K_\D=m_\s^2+\mu_\P^2-2\1{p}_\s\cdot K_\P.\nonumber
}
In the equation, $K_{\D}$ ($K_{\P}$) is the total momentum of the $\D$ ($\P$) system and its invariant mass is denoted as $\mu_{\D(\P)}$, i.e. $K^2_{\D(\P)}=\mu^2_{\D(\P)}$. It is convenient to introduce the mean and the tilded Mandelstam-$t$
\beq\label{eq:tdefs}
t\equiv\frac{\0{q}^2+\1{q}^2}2
\,,\qquad
\di{t}\equiv\frac{\0{q}^2-\1{q}^2}2=-\di{p}\cdot K_\D=\di{p}\cdot K_\P\,.
\eeq
Using $t$ and $\di{t}$ it is natural to express $\Pi$ as
\beq
\label{eq:propsqu}
\Pi=
\frac1{2\di{t}+2i\epsilon}
\left[
\frac1{t-m_\i^2-\di{t}-i\epsilon}-
\frac1{t-m_\i^2+\di{t}+i\epsilon}
\right].
\eeq

For $\di{p}=0$, $\di{t}$ vanishes and $\Pi$ develops a $1/\epsilon$ singularity, where $\epsilon=0^+$ is the unphysical regulator of the Feynman propagator. Therefore the collision function~(\ref{sf2ps}) is non-analytic and singular at $\di{p}=0$. This is the exact same singularity that emerges in the calculation of the cross section in standard scattering theory because the collision function at $\di{p}=0$ is proportional to the cross section up to a regular factor. However, while the singularity is an unphysical divergence in the standard theory, it is harmless in our framework because the transition probability is the integral of the collision function over $\di{p}$---see eq.\,(\ref{eq:MAS3})---and it turns out that the singularity is integrable. The singularity is regulated by the non-vanishing range of integration in $\di{p}$ which is linked to a non-infinite extension in space of the Wigner densities that describe the colliding beams.\,\footnote{\label{fot}We will see in section\,\ref{sec:exc} that the $\di{p}$ singularity is integrable only when the classical momenta of the incoming particles are away from the threshold where the $\i$ propagator can go on-shell on the physical phase-space. Near the threshold, integration over the classical momentum beam spread is also needed for a finite result.}

The $1/\epsilon$ singularity of $\Pi$ at $\di{p}=0$ is proportional to the imaginary part of the terms in the square bracket of eq.\,(\ref{eq:propsqu}), while the real part vanishes when $\di{t}=0$ and therefore it is expected to give a regular and analytic contribution to the collision function. We thus split $\Pi$ as  $\Pi=\Pi_{\rm{r}}+\Pi_{\rm{i}}$ and study the two terms separately. They are given by
\al{\label{eq:Pidecom}
&\Pi_{\rm{r}}=\frac1{2\di{t}+2i\epsilon}
\left[
{\rm{pv}}_\epsilon\hspace{-1pt}(1/
(t-m_\i^2-\di{t}))-
{\rm{pv}}_\epsilon\hspace{-1pt}(1/(t-m_\i^2+\di{t}))
\right],\\
&\Pi_{\rm{i}}=\frac{i\pi}{2\di{t}+2i\epsilon}
\left[
\delta_\epsilon\hspace{-1pt}
(t-m_\i^2-\di{t})+
\delta_\epsilon\hspace{-1pt}(t-m_\i^2+\di{t})
\right],
\label{eq:Pidecomi}
}
where the regulated delta-function and principal value are the familiar expressions
\beq
\frac1{x\mp i\epsilon}=
\frac{x}{x^2+\epsilon^2}
\pm\frac{i\epsilon}{x^2+\epsilon^2}
={\rm{pv}}_\epsilon\hspace{-1pt}(1/x) \pm i\pi\delta_\epsilon\hspace{-1pt}(x)\,.
\eeq

We show now that $\Pi_{\rm{r}}$ gives a regular and analytic contribution to the collision function, by proceeding as follows. The collision function~(\ref{sf2ps}) is an integral over the phase-space, for which we may use a parametrization where $t$ and $\di{t}$ are independent coordinates. If we perform the integral over $t$ at fixed $\di{t}$ we can treat $\di{t}$ as a constant and interpret $\Pi$ as a parameter-dependent distribution with integration variable $t$. All the other factors in the integrand---namely, the decay and scattering amplitudes and their conjugates---provide the test function the $\Pi$ distribution acts on. We notice that they are analytic in $t$ and in $\di{t}$.

By interpreting $\Pi$ as a $\di{t}$-dependent distribution acting on an analytic test function, we show in appendix~\ref{sec:dist} that it can be expanded in series with a finite range of convergence around $\di{t}=0$ by simply considering the geometric series formula for the two terms in the square brackets in eq.\,(\ref{eq:propsqu}). These read
\beq
\frac1{t-m_\i^2\pm\di{t}\pm i\epsilon}=\sum\limits_{n=0}^\infty \frac{(\mp\di{t}\,)^n}{(t-m_\i^2\pm i\epsilon)^{n+1}}\,.
\eeq
Each term of the series is a well-defined distribution for $\epsilon\to0^+$ that can be expressed as in eq.\,(\ref{eq:1/x^n_expansion}) in terms of the Hadamard finite part regularization of $1/x^{n+1}$ and of derivatives of the delta. 
By retaining only the real part we obtain the distribution expansion of $\Pi_{\rm{r}}$
\beq\label{eq:sdpi}
\Pi_{\rm{r}}=
\frac{\di{t}}{\di{t}+i\epsilon}
\sum\limits_{k=0}^\infty\,
{\di{t}\,^{2k}}
{\rm{Pf}}\big[1/(t-m_\i^2)^{2k+2}\big].
\eeq
Only the odd terms of the geometric series survive after taking the real part and the $\Pi_{\rm{r}}$ expansion starts with $\di{t}/(\di{t}+i\epsilon)$, which is regular at $\di{t}=0$ and can be replaced with 1 under the phase-space integral. Therefore each term of the $\Pi_{\rm{r}}$ series contributes to the collision function as the integral of a monomial in $\di{t}=-\di{p}\cdot k_\D$ times the rest of the amplitude and conjugate amplitude that are also analytic functions of $\di{p}$. The contribution of $\Pi_{\rm{r}}$ to the collision function is thus analytic at $\di{p}=0$ as anticipated. 

Analytic terms in the collision function were studied in section~\ref{se:sds} and found to produce a short-distance transition probability---see eq.\,(\ref{eq:MASCL2p})---at the leading order in the classical expansion, plus polynomial corrections that are also short-distance and involve powers of the $\Box$ of the Wigner density. The leading-order transition probability can be interpreted in terms of a cross section that is proportional to the collision function evaluated at $\di{p}=0$. Using the $k=0$ order of the $\Pi_{\rm{r}}$ expansion in eq.\,(\ref{eq:sdpi}) and setting $\di{p}=0$ in the amplitudes it is easy to see that this cross section is equal to the regular integral of the squared amplitudes---like in eq.\,(\ref{eq:scxs})---supplemented by the prescription
\beq\label{eq:simppr}
\frac1{|t-m_\i^2+i\epsilon|^2}\;\longrightarrow\;{\rm{Pf}}\left[\frac1{(t-m_\i^2)^{2}}\right],
\eeq
to handle the ill-defined $t$-channel propagator. This prescription emerges from the physical treatment of the process in spacetime collision theory and it accounts for the contributions to the cross section from $\Pi_{\rm{r}}$. Additional short-distance contributions from $\Pi_{\rm{i}}$ are discussed later and included in the second line of eq.\,(\ref{eq:sigmaan}). The complete expression for the cross section is in eq.\,(\ref{eq:scxsIP}).

The $\Pi_{\rm{i}}$ term~(\ref{eq:Pidecomi}) of the $\Pi$ decomposition delivers both analytic and non-analytic contributions to the collision function. Disentangling the two requires a more explicit expression for the collision function\,(\ref{sf2ps}). Apart from the propagator---which we already squared and included in $\Pi$---the amplitude in figure\,\ref{diag1} is the product of
\beq\label{eq:ampfact}
{\mathcal{M}}_\D(\0{p}_\u;\0{q},\{\hspace{-1pt}k_\D\hspace{-1pt}\})\times
{\mathcal{M}}_\P(\0{q},\0{p}_\s;\{\hspace{-1pt}k_\P\hspace{-1pt}\})\,,
\eeq
and similarly for the conjugate amplitude, but with dotted momenta. The phase-space integration variables are the momenta $\{\hspace{-1pt}k_\D\hspace{-1pt}\}$ and $\{\hspace{-1pt}k_\P\hspace{-1pt}\}$ of the $\D$ and $\P$ multi-particle states. Since $\Pi$ depends only on the total $\D$ and $\P$ momenta $K_\D$ and $K_\P$, it is convenient to introduce those as integration variables by writing
\al{
&\prod\limits_{i\in\D}
\frac{d^3\hspace{-2pt}\3{k}_i}{(2\pi)^32k_i^0}
=\frac{d^4\hspace{-2pt}K_\D}{(2\pi)^4}\,\d{\rm{Ph}}_\D\hspace{-1pt}(K_\D)
=\frac{d\mu_\D^2}{2\pi}
\frac{d^3\hspace{-2pt}\3{K}_\D}{(2\pi)^32{K}^0_\D}\,\d{\rm{Ph}}_\D\hspace{-1pt}(K_\D)
\,,\\
&\prod\limits_{i\in\P}
\frac{d^3\hspace{-2pt}\3{k}_i}{(2\pi)^32k_i^0}
=\frac{d^4\hspace{-2pt}K_\P}{(2\pi)^4}\,\d{\rm{Ph}}_\P\hspace{-1pt}(K_\P)
=\frac{d\mu_\P^2}{2\pi}
\frac{d^3\hspace{-2pt}\3{K}_\P}{(2\pi)^32{K}^0_\P}\,\d{\rm{Ph}}_\P\hspace{-1pt}(K_\P)
\,.\nonumber
}
We also introduced the variables $\mu_\D$ and $\mu_\P$ that are the invariant masses of $K_\D$ and $K_\P$. The total phase-space volume element is thus cast in the factorised form
\beq
\d{\rm{Ph}}(p_{\rm{tot}})=
\frac{d\mu_\P^2}{2\pi}\frac{d\mu_\D^2}{2\pi} \d{\rm{Ph}}_{2}(p_{\rm{tot}})\d{\rm{Ph}}_\D\hspace{-1pt}(K_\D)\d{\rm{Ph}}_\P\hspace{-1pt}(K_\P)\,,
\eeq
where $\d{\rm{Ph}}_{2}$ is the 2-body phase space for particles with mass $\mu_\D$ and $\mu_\P$:
\beq
\d{\rm{Ph}}_{2}(p_{\rm{tot}})=
(2\pi)^4\delta^4(K_\D\hspace{-2pt}+\hspace{-2pt}K_\P\hspace{-2pt}-\hspace{-2pt}p_{\rm{tot}})
\frac{d^3\hspace{-2pt}\3{K}_\D}{(2\pi)^32{K}^0_\D}\frac{d^3\hspace{-2pt}\3{K}_\P}{(2\pi)^32{K}^0_\P}\,.
\eeq

The integrals over the $\D$ and the $\P$ phase-spaces can now be combined with the corresponding amplitudes~(\ref{eq:ampfact}) and conjugate amplitudes and collected into Lorentz scalar functions $\mathfrak{D}$ and $\mathfrak{S}$. These are given by
\al{
\label{eq:PD}
&\mathfrak{D}(\di{t},t,-\di{p}^2,\mu_\D^2)
=\In{4}
\d{\rm{Ph}}_\D\hspace{-1pt}(K_\D)
{\mathcal{M}}_\D(\0{p}_\u;\0{q},\{\hspace{-1pt}k_\D\hspace{-1pt}\})
{\mathcal{M}}_\D^*(\1{p}_\u;\1{q},\{\hspace{-1pt}k_\D\hspace{-1pt}\})\,,\\
&\mathfrak{S}(\di{t},t,-\di{p}^2,\mu_\P^2)
=\In{4}
\d{\rm{Ph}}_\P\hspace{-1pt}(K_\P)
{\mathcal{M}}_\P(\0{q},\0{p}_\s;\{\hspace{-1pt}k_\P\hspace{-1pt}\})
{\mathcal{M}}_\P^*(\1{q},\1{p}_\s;\{\hspace{-1pt}k_\P\hspace{-1pt}\})\,.
\nonumber
}
Lorentz symmetry constrains the functional dependence of $\mathfrak{D}$ and $\mathfrak{S}$. The function $\mathfrak{D}$ depends on $p_\u$, $\di{p}$ and $K_\D$---clearly $\zu{q}$ is not an independent variable because of momentum conservation---but $p_\u$ and $\di{p}$ are orthogonal and therefore the only independent invariants are $p_\u\hspace{-1pt}\cdot\hspace{-1pt}  K_\D$, $\di{p}\hspace{-1pt}\cdot\hspace{-2pt} K_\D$, $\di{p}^2$ and $K_\D^2\hspace{-2pt}=\hspace{-2pt}\mu_D^2$. The first two invariants are equivalent to $\di{t}$ and ${t}$:
\beq\label{eq:ttt}
\di{t}\hspace{-2pt}=\hspace{-2pt}-\di{p}\cdot K_\D\,,\qquad
t=m_\u^2+\mu_\D^2-2p_\u\cdot K_\D\,.
\eeq
One can similarly work out the arguments of the function $\mathfrak{S}$.

Collecting all factors gives an expression for the collision function
\beq\label{eq:sfexp}
{\mathfrak{C}}(S,-\di{p}^2)\hspace{-1pt}=\hspace{-1pt}
\In{4}\frac{d\mu_\P^2}{2\pi}
\In{4}
\frac{d\mu_\D^2}{2\pi} \In{5}\d{\rm{Ph}}_{2}(p_{\rm{tot}})\,
\Pi(\di{t},t)\,\mathfrak{D}\hspace{-1pt}\mathfrak{S}\hspace{-1pt}
(\di{t},t,-\di{p}^2)
\,,
\eeq
where $\mathfrak{D}\hspace{-1pt}\mathfrak{S}$ denotes the product of $\mathfrak{D}$ with $\mathfrak{S}$. The dependence of $\mathfrak{D}\hspace{-1pt}\mathfrak{S}$ on $\mu_\D$ and $\mu_\P$ is not reported for brevity.

The 2-body phase-space can be parametrized by the variables $t$ and $\di{t}$ (see eq.~(\ref{phsttt})) and eq.\,(\ref{eq:sfexp}) cast into an explicit representation of ${\mathfrak{C}}$ as a $t$-$\di{t}$ integral of $\Pi$ times an analytic function, namely $\mathfrak{D}\hspace{-1pt}\mathfrak{S}$. This allows for the series expansion of $\Pi$ in $\di{t}$ as previously explained. By this expansion we proved that the $\Pi_{\rm{r}}$ component of $\Pi$ defined in~(\ref{eq:Pidecom}) gives an analytic contribution to the collision function. Possible non-analyticities emerge from $\Pi_{\rm{i}}$~(\ref{eq:Pidecomi}) due to the singular $1/(\di{t}+i\epsilon)$ prefactor, but only when the rest of the integrand (i.e., $\mathfrak{D}\hspace{-1pt}\mathfrak{S}$) does not contain a positive power of $\di{t}$ that cancels the singularity. Non-analyticities can thus only emerge from the first term of the Taylor series expansion of $\mathfrak{D}\hspace{-1pt}\mathfrak{S}$ around $\di{t}=0$ while if we subtract from $\mathfrak{D}\hspace{-1pt}\mathfrak{S}$ its value at $\di{t}=0$ we get analytic terms. The analytic component of the collision function is thus the sum of the $\Pi_{\rm{r}}$ contribution, plus the one of $\Pi_{\rm{i}}$ acting on the subtracted $\mathfrak{D}\hspace{-1pt}\mathfrak{S}$ function:
\al{
\label{eq:sigmaan}
{\mathfrak{C}}_{a}(S,-\di{p}^2)\hspace{-1pt}=\hspace{-1pt}
\In{4}\frac{d\mu_\P^2}{2\pi}
\In{4}
\frac{d\mu_\D^2}{2\pi} \In{5}\d{\rm{Ph}}_{2}(p_{\rm{tot}}&)\bigg\{
\Pi_{\rm{r}}(\di{t},t)\,\mathfrak{D}\hspace{-1pt}\mathfrak{S}\hspace{-1pt}
(\di{t},t,-\di{p}^2)
\\
&\hspace{-5pt}+
\Pi_{\rm{i}}(\di{t},t)\left[\mathfrak{D}\hspace{-1pt}\mathfrak{S}\hspace{-1pt}
(\di{t},t,-\di{p}^2)-
\mathfrak{D}\hspace{-1pt}\mathfrak{S}\hspace{-1pt}
(0,t,-\di{p}^2)\right]
\bigg\}.\nonumber
}

Analytic terms in the collision function correspond to short-distance terms in the transition probability and at the leading order in the classical expansion they are the product of the Wigner density at the same spacetime point, times a factor that can be interpreted as a cross section. This defines the cross section for the unstable particle
\beq\label{eq:scxsIP}
\sigma(\u\,\s\to\D\,\P)\equiv
{\scriptstyle{\frac1{4\sqrt{(P_\u\cdot P_\s)^2-m_\u^2m_\s^2}}}}\;{\mathfrak{C}}_a(S,0),
\eeq
where ${\mathfrak{C}}_a(S,0)$ is given by eq.\,(\ref{eq:sigmaan}) taking the $\di{p}\to0$ limit of the integrand. The contribution on the first line, from $\Pi_{\rm{r}}$, is merely the regular cross section formula supplemented with the simple prescription in eq.\,(\ref{eq:simppr}) as previously discussed. The contribution from $\Pi_{\rm{i}}$ can be worked out easily and it is proportional to the derivative of $\mathfrak{D}\hspace{-1pt}\mathfrak{S}$ with respect to $\di{t}$ evaluated at $\di{p}=0$ and $t=m_\i^2$.

The remainder of the $\mathfrak{D}\hspace{-1pt}\mathfrak{S}$ subtraction gives
\beq\label{eq:nan}
{\mathfrak{C}}_{\slashed{a}}(S,-\di{p}^2)\hspace{-1pt}=\hspace{-1pt}
\In{4}\frac{d\mu_\P^2}{2\pi}
\In{4}
\frac{d\mu_\D^2}{2\pi} \In{5}\d{\rm{Ph}}_{2}(p_{\rm{tot}})\,
\Pi_{\rm{i}}(\di{t},t)\,\mathfrak{D}\hspace{-1pt}\mathfrak{S}\hspace{-1pt}
(0,t,-\di{p}^2)
\,.
\eeq
We now show that ${\mathfrak{C}}_{\slashed{a}}$ is non-analytic and singular at $\di{p}=0$ and we assess the structure of the singularity. We will find that the singularity is integrable in $\di{p}$ (see however footnote\,\ref{fot}) and therefore it produces a well-defined transition probability.  

Eq.\,(\ref{eq:nan}) can be simplified by noticing that the collision function depends only on $\di{p}^2$ and on $S$ and thus it is invariant if we change sign to the $\di{p}$ four-vector keeping fixed the classical momenta---and the mean momenta, since their relation with the classical ones only involves $\di{p}^2$---that describe the incoming particles. The invariance of ${\mathfrak{C}}$ under $\di{p}\to-\di{p}$ is a completely general consequence of Lorentz symmetry that holds at finite $\epsilon$ and is unrelated with the possible presence of non-analyticities for $\epsilon\to0^+$. It can be also proven exploiting the existence of a Lorentz transformation that changes sign to $\di{p}$ while preserving the classical (or mean) momenta of the incoming particles. This transformation exists because $\di{p}$ is space-like and it is orthogonal to the classical momenta. In the centre-of-mass frame it is simply a rotation around the direction of the incoming particles. Changing variables in the phase-space integral---either in the one of eq.\,(\ref{eq:nan}) or in the general formula\,(\ref{sf2ps})---is equivalent to change the sign of $\di{p}$ in the integrand.

If we symmetrize the integrand of eq.\,(\ref{eq:nan}) with respect to $\di{p}\to-\di{p}$ only the delta-function component of the $1/(\di{t}+i\epsilon)$ factor of $\Pi_{\rm{i}}$~(\ref{eq:Pidecomi}) survives and we obtain
\al{
{\mathfrak{C}}_{\slashed{a}}(S,-\di{p}^2)\hspace{-1pt}=&\hspace{-1pt}
\In{4}\frac{d\mu_\P^2}{2\pi}
\In{4}
\frac{d\mu_\D^2}{2\pi} \\
&\In{5}\d{\rm{Ph}}_{2}(p_{\rm{tot}})\bigg\{\frac{\pi^2}2\delta_\epsilon(\di{t})
\left[
\delta_\epsilon\hspace{-1pt}
(t-m_\i^2-\di{t})+
\delta_\epsilon\hspace{-1pt}(t-m_\i^2+\di{t})
\right]
\,\mathfrak{D}\hspace{-1pt}\mathfrak{S}\hspace{-1pt}
(0,t,-\di{p}^2)\bigg\}
\,.\nonumber
}
Taking the $\epsilon\to0^+$ limit we get
\beq\label{eq:Snan}
{\mathfrak{C}}_{\slashed{a}}(S,-\di{p}^2)\hspace{-1pt}=\hspace{-1pt}
\In{4}\frac{d\mu_\P^2}{2\pi}
\In{4}
\frac{d\mu_\D^2}{2\pi}
\mathfrak{D}\hspace{-1pt}\mathfrak{S}\hspace{-1pt}
(0,m_\i^2,-\di{p}^2)
\In{4}\d{\rm{Ph}}_{2}(p_{\rm{tot}})\,
\pi^2\delta(\di{t})\,\delta(t-m_\i^2)\,.
\eeq
This shows that the non-analytic part of ${\mathfrak{C}}$ arises from the double-delta component of the squared propagator. We can obtain ${\mathfrak{C}}_{\slashed{a}}$ directly from the master formula~(\ref{sf2ps}) by replacing
\beq\label{eq:Pi_na}
\Pi\;\longrightarrow\;\Pi_{\slashed{a}}\equiv\pi^2\delta(\di{t})\,\delta(t-m_\i^2)\,.
\eeq

The result is compatible with the intuition that the non-analyticity is associated with the kinematic configurations where the $t$-channel momentum reaches the mass-shell of the intermediate particle $\i$. More precisely, it emerges from the configuration where $t=m_\i^2$ and $\di{t}=0$ which means---see eq.\,(\ref{eq:tdefs})---that both the circled and dotted $t$-channel momenta $\0{q}$ and $\1{q}$ reach the mass-shell simultaneously, i.e. at the same point of the phase-space. 

A more explicit formula for ${\mathfrak{C}}_{\slashed{a}}$ can be obtained straightforwardly by integrating the delta-functions over the two-body phase-space---see section~\ref{sec:exc}. This gives
\beq\label{eq:sf_nonanalitic}
{\mathfrak{C}}_{\slashed{a}}(S,-\di{p}^2)\hspace{-1pt}=
\frac1{\sqrt{-\di{p}^2}}
\hspace{1pt}
\In{4}
\frac{d\mu_\P^2}{2\pi}
\In{4}
\frac{d\mu_\D^2}{2\pi}
\mathfrak{D}\hspace{-1pt}\mathfrak{S}\hspace{-1pt}
(0,m_\i^2,-\di{p}^2) \frac{1}{4\sqrt{\Lambda_F}} \frac{\Theta(\di{p}^2 \hspace{-2pt}+\hspace{-2pt} \alpha)}{\sqrt{\di{p}^2 \hspace{-2pt}+\hspace{-2pt} \alpha}}
\,,
\eeq
where $\Lambda_F$ (see table\,\ref{tab:2bpp}) and $\alpha$~(\ref{eq:alpha}) are functions of $s$ and of the masses. The condition $\alpha \geq -\di{p}^2$ enforced by the theta function is required to ensure that the circled and dotted momenta can go on-shell simultaneously in the kinematically allowed phase-space. If this configuration is not allowed, the non-analyticities are absent as discussed above.

The explicit expression shows that ${\mathfrak{C}}_{\slashed{a}}$ is non-analytic and singular at $\di{p}=0$. The transition probability is the integral of ${\mathfrak{C}}$ over $\di{p}$ and we established in eq.\,(\ref{eq:vintpt}) that the integration measure is two-dimensional. Therefore the $1/\sqrt{-\di{p}^2}$ singularity is integrable and the transition probability is well-defined. Other interesting features of eq.\,(\ref{eq:sf_nonanalitic}), including the presence of a singularity at $\di{p}^2=-\alpha$, are discussed in section\,\ref{sec:exc}.

\subsection{Long distance transition probability}
\label{sec:ldt}

In this section we study the contributions to the transition probability that arise from the non-analyticities in
${\mathfrak{C}}$. We will find that they feature a long distance structure, namely they are proportional to the product of the Wigner densities (or derivatives thereof, beyond the leading order in the classical expansion) evaluated at faraway spacetime points. The two points correspond to the position of the $\u$ decay process, which produces $\i$, and to the one of the $\i\,\s$ scattering. They are connected by the trajectory  of the $\i$ particle.

The leading-order result can be extracted by studying the behaviour at small $\di{p}^2$ of the $\Pi_{\rm{i}}$ part~(\ref{eq:Pidecomi}) of the squared propagator. At small $\di{p}$ the mean momenta are equal to the classical momenta and therefore
\beq\label{eq:Qdef}
t=m_\u^2+\mu_\D^2-2p_\u\cdot K_\D\simeq(P_\u-K_\D)^2=Q^2\,,
\eeq
where we defined $Q=P_\u- K_\D$. Furthermore, since $P_\u\cdot\di{p}=0$, we can express
\beq
\di{t}=-\di{p}\cdot K_\D=    \di{p}\cdot(P_\u- K_\D)=\di{p}\cdot Q\,,
\eeq
and approximate $\Pi_{\rm{i}}$ as
\beq
\Pi_{\rm{i}}\simeq\frac{i\pi}{\di{p}\cdot Q+i\epsilon}\delta(Q^2-m_\i^2).
\eeq
Notice that it is legitimate to use an approximation of $t$ inside the delta because it acts as a  distribution on an analytic test function. 

The variable $Q$ is the classical limit of the $t$-channel momenta $\0{q}$ and $\1{q}$---see figure\,\ref{diag1}---of the intermediate particle $\i$. Since $\Pi_{\rm{i}}$ sets it on-shell, it can be interpreted as the four-momentum of a physically-propagating $\i$ particle. Actually $Q$ is the physical momentum only if $Q^0$ is positive, in which case it describes propagation from the $\D$ to the $\P$ blobs of figure\,\ref{diag1}. Otherwise the relevant momentum is $-Q$, describing the propagation from $\P$ to $\D$. We can separate these two configurations by treating separately the $Q^0\gtrless0$ regions of the phase-space. We focus on $Q^0>0$ and comment later on the treatment of the other region, which is essentially identical.

The delta-function from $\Pi_{\rm{i}}$ combines with the phase-space volume element by the standard trick\,\footnote{The theta-function is present because we are at $Q^0>0$. With an abuse of notation, on the left hand side of the equation $Q$ should be interpreted as $Q=P_\u- K_\D$, where $K_\D = \sum k_\D$.}
\beq\label{eq:st}
\Theta(Q^0)\delta(Q^2-m_\i^2)\d{\rm{Ph}}(P_{\rm{tot}})=\frac1{(2\pi)^4}
\frac{d^3\hspace{-1pt}\3{Q}}{2E\hspace{-1pt}(\3{Q})}
\d{\rm{Ph}}_\D\hspace{-1pt}(P_\u-Q)\,
\d{\rm{Ph}}_\P\hspace{-1pt}(Q+P_\s),
\eeq
where on the right side of the equation $Q$ is a physical particle momentum with temporal component $Q^0=E\hspace{-1pt}(\3{Q})=({\3{Q}^2+m_\i^2})^{1/2}$. In the equation, $\d{\rm{Ph}}_\D$ and $\d{\rm{Ph}}_\P$ are the phase-space volume elements of the $\D$ and $\P$ multi-particle (or single-particle) states. Originally the total momentum on the phase-space was $p_{\rm{tot}}$, however it has been replaced by the total classical momentum $P_{\rm{tot}}=P_\u+P_\s$ because the two are equal in the classical limit.

The factorisation of the phase-space combines naturally with the factorized form of the $t$-channel amplitude, which is the product of the $\D$ and the $\P$ amplitudes as in eq.~(\ref{eq:ampfact}) times the product of the conjugate amplitudes, with dotted momenta. In the classical limit all the momenta including $\0{q}$ (and $\1{q}$, in the conjugate amplitude) can be replaced with the classical momenta and since $Q$ is on-shell we recover the physical amplitudes for the decay in eq.\,(\ref{eq:decc}) and the scattering process in eq.\,(\ref{eq:prod}). 

The squared ${\mathcal{M}}_\D$ combines with the integral over $\d{\rm{Ph}}_\D\hspace{-1pt}$ to reconstruct the differential decay rate of the $\u\to\i\,\D$ process
\beq\label{eq:partialwidth}
\frac1{(2\pi)^32E(\3{Q})}\In{4}\d{\rm{Ph}}_\D\hspace{-1pt}(P_\u-Q)\,|{\mathcal{M}}_\D(P_\u;Q,\{\hspace{-1pt}k_\D\hspace{-1pt}\})|^2=2m_\u\frac{d^3\Gamma_\D\hspace{-1pt}}{d^3\3{Q}}\,,
\eeq
where we also included $1/E(\3{Q})$ and the other factors of eq.\,(\ref{eq:st}) that correspond to the relativistic integration measure on the $\i$ particle momentum. The squared ${\mathcal{M}}_\P$ gives the $\i\,\s\to\P$ cross section:
\beq\label{eq:productionxs}
\In{4}\d{\rm{Ph}}_\P\hspace{-1pt}(Q+P_\s)\,|{\mathcal{M}}_\P(Q,P_\s,\{\hspace{-1pt}k_\P\hspace{-1pt}\})|^2=
{\scriptstyle{4\sqrt{(Q\cdot P_\s)^2-m_\i^2m_\s^2}}}\;\sigma_\P\hspace{-1pt}\,.
\eeq
Collecting all the factors we obtain the contribution from $\Pi_{\rm{i}}$ to the collision function at the leading order in the classical expansion
\beq\label{eq:losl}
{\mathfrak{C}}_{{\rm{i}}}^0=\,
\In{5}d^3\hspace{-1pt}\3{Q}
\frac{d^3\hspace{-1pt}\Gamma_\D}{d^3\hspace{-1pt}\3{Q}}\,
\frac{i\,m_\u}{\di{p}\cdot Q+i\epsilon}\,\left[{\scriptstyle{4\sqrt{(Q\cdot P_\s)^2-m_\i^2m_\s^2}}}\,\sigma_\P\right]\,.
\eeq

The last step is to insert ${\mathfrak{C}}_{{\rm{i}}}^0$ in eq.\,(\ref{eq:MAS3}) in order to obtain the transition probability. Given that we would like to recover the Wigner densities that are the Fourier transform of the density matrices with respect to $\di{p}$, it is convenient to employ the Fourier representation for the $\di{p}$-dependent factor in the expression for  ${\mathfrak{C}}_{{\rm{i}}}^0$, namely
\beq\label{eq:distour}
\frac{i}{\di{p}\cdot Q+i\epsilon}
=\frac1{m_\i}\,\BIn{3}{-\infty}{0}d\tau\,
e^{-i{\tau} [\di{p}\cdot Q +i\epsilon]/m_\i }=\frac1{m_\i}\,
\BIn{3}{-\infty}{0}d\tau\,
e^{-i \tau\, \di{p}_\u\cdot Q/m_\i}e^{\epsilon\tau/m_\i}\,.
\eeq
For convenience we expressed $\di{p}$ as $\di{p}_\u$ and we used $m_\i$ to normalize the integration variable $\tau$ to have the dimension of time. The imaginary exponential combines with the $\di{p}_\u\cdot{x}$ exponential that is present in eq.\,(\ref{eq:MAS3}) while the $\epsilon=0^+$ convergence factor can be safely dropped. We obtain the transition probability
\al{
\label{eq:ldsp}
&p_{\rm{i}}^0=
\In{3}d^4\hspace{-1pt}x
\,\In{3}d^3\hspace{-1pt}\3{Q}\,
\In{3}{d^3\hspace{-2pt}\3{P}_\s}\Bigg\{\\
&\qquad\qquad\;
\In{3}{d^3\hspace{-2pt}\3{P}_\u}\hspace{-2pt}
\,\BIn{3}{-\infty}{0}d\tau\,\frac{m_\u Q^0}{m_\i P_\u^0}
\frac{d^3\hspace{-1pt}\Gamma_\D}{d^3\hspace{-1pt}\3{Q}}
\underbrace{\left[
\frac1{(2\pi)^6}\frac1{2P^0_\u}
\In{3}
{d^3\hspace{-1pt}\di{\3{p}}_\u}
\frac{p^2_\u}{m_\u^2}\rho_\u
e^{-i\di{p}_\u\cdot{(x+\tau Q/m_\i)}}
\right]}_{\displaystyle{\mathcal{P}}_\u(x+\tau Q/m_\i,\3{P}_\u)}
\nonumber\\
&
\qquad\;\;\;\;
\times\underbrace{\left[
\frac1{(2\pi)^6}\frac1{2P^0_\s}
\In{3}
{d^3\hspace{-1pt}\di{\3{p}}_\s}
\frac{p^2_\s}{m^2_\s}\rho_\s
e^{-i\di{p}_\s\cdot x}
\right]}_{\displaystyle{\mathcal{P}}_\s(x,\3{P}_\s)}
\bigg[\di{v}_{\rm{rel}}\,\sigma_\P\bigg]
\Bigg\},
\qquad\; \di{v}_{\rm{rel}}={\scriptstyle \frac{\sqrt{(Q\cdot P_\s)^2-m_\i^2m_\s^2}}{Q^0P^0_\s}}
\,.
\nonumber
}

This formula cannot be interpreted classically as the result of short-distance scattering between $\u$ and $\s$ because the corresponding Wigner densities are evaluated at different spacetime positions. However it can be interpreted as the short-distance scattering of classical particles $\i$ and $\s$, where the $\i$ particle emerges from the short-distance (localized in space and instantaneous in time) decay of the unstable $\u$. The scattering takes place at $x$---which is where ${\mathcal{P}}_\s$ is evaluated---and its probability is $\sigma_\P$ times the familiar velocity factor. The space-momentum density for the $\i$ particle, with momentum $\3{Q}$ and position $x$, is effectively provided by the second line of the equation. It corresponds to the density of $\i$ particles that is generated by the decay of the $\u$ particle beam and it is obtained by an integral over the $\i$ particle trajectory
\beq
x^\mu+\frac{\tau}{m_\i}Q^\mu,
\eeq
parametrised by the proper time $\tau$. The point on the trajectory is the spacetime position of the decay of $\u$, which consistently happens earlier than the scattering because $\tau<0$. The probability density for the particle $\i$ at $x$ is thus given by the density of $\u$ particles at the decay point, times the differential decay width, times the infinitesimal time interval measured in the rest frame of the decaying particle $\u$. Since $\tau$ is the proper time of $\i$, the infinitesimal interval is $m_\u Q^0/(m_\i P_\u^0)\cdot d\tau$ as in the equation.

So far we discussed the contribution to the collision function from the region of the phase-space integral where $Q^0=P_\u^0-K_\D^0>0$. In the other region, the relevant physical momentum has the opposite sign and it can be written as $Q=K_\D-P_\u=P_\s-K_\P$ by momentum conservation. Therefore the previous derivation applies identically provided we exchange the role of $\u$ with $\s$ and of $\D$ with $\P$. The result is a second long-distance contribution to the transition probability that corresponds to the $\s$ decay followed by the scattering with $\u$ of its decay product $\i$. Notice that the $\s$ particle must be also unstable for this second contribution to appear because solving the on-shell condition is equivalent to solving the kinematics for the $\s\to\i\,\P$ decay process. If $\s$ is instead stable, eq.\,(\ref{eq:ldsp}) is the only long-distance component of the transition probability at the leading order in the classical expansion.

Beyond the leading order, the $\Pi_{\rm{i}}$ contribution to the collision function also delivers analytic terms and the genuine non-analyticities stem from the $\Pi_{\slashed{a}}$ part of the squared propagator that we identified in eq.\,(\ref{eq:Pi_na}). We now study its contribution to the collision function exactly, before taking the classical limit. We define $q$ as the mean momentum that flows in the $t$-channel
\beq\label{eq:qdef}
q=p_\u-K_\D=K_\P-p_\s\,,
\eeq
and we use it to factorize the phase-space as
\beq
\d{\rm{Ph}}(p_{\rm{tot}})=
\frac{d^4\hspace{-1pt}{q}}{(2\pi)^4}\,
\d{\rm{Ph}}_\D\hspace{-1pt}(p_\u-q)\,
\d{\rm{Ph}}_\P\hspace{-1pt}(q+p_\s).
\eeq
The squared propagator depends only on $q$, specifically 
\beq
\Pi_{\slashed{a}}=\pi^2\delta(\di{t})\,\delta(t-m_\i^2)=\pi^2\delta(\di{p}\cdot{q})\,\delta(q^2-m_\i^2+\di{p}^2/4)\,,
\eeq
where we used eq.\,(\ref{eq:ttt}) and the identities $p_\u^2\hspace{-1pt}=\hspace{-1pt}m_\u^2\hspace{-1pt}-\hspace{-1pt}\di{p}^2\hspace{-1pt}/4$ and $p_\u\cdot\di{p}=0$ in order to express $t=q^2\hspace{-1pt}+\hspace{-1pt}\di{p}^2\hspace{-1pt}/4$ and $\di{t}=\di{p}\hspace{-1pt}\cdot\hspace{-1pt}{q}$. We can thus perform the integral over the other phase-space variables $\{\hspace{-1pt}k_\P\hspace{-1pt}\}$ and $\{\hspace{-1pt}k_\D\hspace{-1pt}\}$ that characterise the $\P$ and $\D$ multi-particles states. This gives two factors that collect the phase-space  and the amplitudes of the scattering and of the decay process, respectively
\al{
\label{eq:2fac}
\In{4}
\d{\rm{Ph}}_\P\hspace{-1pt}(q\hspace{-1pt}+\hspace{-1pt}{p}_\s)
{\mathcal{M}}_\P\hspace{-1pt}(q+\di{p}/2,{p}_\s\hspace{-1pt}-\hspace{-1pt}\di{p}/2;\{\hspace{-1pt}k_\P\hspace{-1pt}\})
{\mathcal{M}}_\P^*\hspace{-1pt}(q\hspace{-1pt}-\hspace{-1pt}\di{p}/2,{p}_\s\hspace{-1pt}+\hspace{-1pt}\di{p}/2;\{\hspace{-1pt}k_\P\hspace{-1pt}\})\\
={\mathfrak{S}}
(\di{p}\hspace{-1pt}\cdot\hspace{-1pt}{q}\,,q^2\hspace{-2pt}+\hspace{-2pt}\di{p}^2\hspace{-1pt}/\hspace{-1pt}4,\,-\di{p}^2,\,(p_\s\hspace{-1pt}+\hspace{-1pt}q)^2)\times
\mathds{1}\hspace{-1pt}\{q\hspace{-1pt}+\hspace{-1pt}{p}_\s\in I_+\}
,\nonumber\\
\In{4}
\d{\rm{Ph}}_\D\hspace{-1pt}({p}_\u\hspace{-1pt}-\hspace{-1pt}q)
{\mathcal{M}}_\D({p}_\u\hspace{-1pt}+\hspace{-1pt}\di{p}/2;q\hspace{-1pt}+\hspace{-1pt}\di{p}/2,\{\hspace{-1pt}k_\D\hspace{-1pt}\})
{\mathcal{M}}_\D^*\hspace{-1pt}({p}_\u\hspace{-1pt}-\hspace{-1pt}\di{p}/2;q\hspace{-1pt}-\hspace{-1pt}\di{p}/2,\{\hspace{-1pt}k_\D\hspace{-1pt}\})\nonumber\\
={\mathfrak{D}}
(\di{p}\hspace{-1pt}\cdot\hspace{-1pt}{q}\,,q^2\hspace{-1pt}+\hspace{-1pt}\di{p}^2\hspace{-1pt}/\hspace{-1pt}4,\,-\di{p}^2,\,(p_\u\hspace{-1pt}-\hspace{-1pt}q)^2)\times
\mathds{1}\hspace{-1pt}\{p_\u\hspace{-1pt}-\hspace{-1pt}q\in I_+\}
.\nonumber
}

The factors can be expressed in terms of the functions ${\mathfrak{S}}$ and ${\mathfrak{D}}$ defined in eq.\,(\ref{eq:PD}), taking however into account that these functions are defined with total $K_\P$ and $K_\D$ momenta that belong to the interior of the positive light-cone $I_+$---i.e., they have positive energy and $K_{\P,\D}^2\hspace{-2pt}=\hspace{-2pt}\mu_{\P,\D}^2\hspace{-2pt}>\hspace{-2pt}0$---while $q\hspace{-1pt}+\hspace{-1pt}p_\s$ and $p_\u\hspace{-1pt}-\hspace{-1pt}q$ are generic four-vectors. The phase-space integrals are non-zero only if the vectors belong to $I_+$, which gives rise to the indicator functions ``$\mathds{1}$'' in eq.\,(\ref{eq:2fac}). Given a generic time-like vector $v$, the condition $v^0\hspace{-1pt}>\hspace{-1pt}0$ is equivalent to the condition that $v$ has positive scalar product with an arbitrary time-like and positive energy vector $w\in I_+$. By choosing $w$ to be $p_\s$ or $p_\u$ we can thus write the indicator functions as
\al{
&\mathds{1}\hspace{-1pt}\{q\hspace{-1pt}+\hspace{-1pt}{p}_\s\in I_+\}=\Theta((q\hspace{-1pt}+\hspace{-1pt}{p}_\s)^2)\,\Theta({p}_\s^2\hspace{-1pt}+\hspace{-1pt}
q\hspace{-1pt}\cdot\hspace{-1pt}{p}_\s),\nonumber\\
&\mathds{1}\hspace{-1pt}\{p_\u\hspace{-1pt}-\hspace{-1pt}q\in I_+\}=\Theta((p_\u\hspace{-1pt}-\hspace{-1pt}q)^2)\,\Theta(p_\u^2\hspace{-1pt}-\hspace{-1pt}
q\hspace{-1pt}\cdot\hspace{-1pt}{p}_\u),
}
This makes manifest that they are Lorentz-invariant and that they depend on the same invariants as the $\mathfrak{S}$ and $\mathfrak{D}$ terms in eq.\,(\ref{eq:2fac}).

The delta-functions in $\Pi_{\slashed{a}}$ fix $q^2\hspace{-1pt}=\hspace{-1pt}m_\i^2\hspace{-1pt}-\hspace{-1pt}\di{p}^2\hspace{-1pt}/\hspace{-1pt}4$ and $\di{p}\hspace{-1pt}\cdot\hspace{-1pt}q\hspace{-1pt}=\hspace{-1pt}0$. The residual independent invariants (recall $p_{\u,\s}^2\hspace{-3pt}=\hspace{-1pt}m_{\u,\s}^2\hspace{-2pt}-\hspace{-1pt}\di{p}^2\hspace{-1pt}/\hspace{-1pt}4$) are $\di{p}^2$, $p_\u\hspace{-2pt}\cdot\hspace{-1pt}q$ and $p_\s\hspace{-2pt}\cdot\hspace{-1pt}q$. Therefore the collision function takes the form
\beq\label{eq:Snan1}
{\mathfrak{C}}_{\slashed{a}}(S,-\di{p}^2)\hspace{-1pt}=\hspace{-1pt}\pi^2\,
\In{5}\frac{d^4\hspace{-1pt}{q}}{(2\pi)^4}\,
\delta(\di{p}\cdot{q})\,\delta(q^2\hspace{-2pt}-\hspace{-2pt}m_\i^2\hspace{-2pt}+\hspace{-2pt}\di{p}^2\hspace{-1pt}/\hspace{-1pt}4)\,
f(-\di{p}^2,p_\u\hspace{-2pt}\cdot\hspace{-1pt}q,p_\s\hspace{-2pt}\cdot\hspace{-1pt}q)
\;,
\eeq
where $f$ is the product of the two factors in eq.\,(\ref{eq:2fac}). This same expression could have been also obtained starting from eq.\,(\ref{eq:Snan}).

Eq.\,(\ref{eq:Snan1}) is not compatible with the physical intuition that the non-analyticities of the collision function are due to the on-shell propagation. The interpretation is respected only at the leading order in the classical expansion, where the second delta function sets indeed $q^2\hspace{-1pt}=\hspace{-1pt}m_\i^2$, but it is seemingly violated at higher order. However, it is possible to express ${\mathfrak{C}}_{\slashed{a}}$ as the integral over an on-shell momentum $Q$ by proceeding as follows. Consider changing variables in the $q$ integral with any of these two transformations 
\beq
q=Q\pm \di{p}/2\,.
\eeq
Notice that $p_{\u,\s}\hspace{-2pt}\cdot\hspace{-1pt}q=p_{\u,\s}\hspace{-2pt}\cdot\hspace{-1pt}Q$ because $\di{p}$ is orthogonal to $p_{\u,\s}$, therefore the arguments of the function $f$ are invariant under the change of variables. The two delta-functions give
\beq
\delta(\di{p}\cdot{q})\,\delta(q^2\hspace{-2pt}-\hspace{-2pt}m_\i^2\hspace{-2pt}+\hspace{-2pt}\di{p}^2\hspace{-1pt}/\hspace{-1pt}4)=\delta(\di{p}\cdot{Q}\hspace{-1pt}\pm\hspace{-1pt}\di{p}^2\hspace{-1pt}/\hspace{-1pt}2)\,\delta(Q^2\hspace{-2pt}-\hspace{-2pt}m_\i^2\hspace{-1pt}),
\eeq
such that $Q$ is on-shell for both sign choices. Operating the two transformations gives two equivalent expressions and  ${\mathfrak{C}}_{\slashed{a}}$ can be written as the average of the two
\beq
{\mathfrak{C}}_{\slashed{a}}(S,-\di{p}^2)\hspace{-1pt}=\hspace{-1pt}\frac{\pi^2}2\,
\In{5}\frac{d^3\hspace{-1pt}\3{Q}}{(2\pi)^42E(\3{Q})}\,\big[
\delta(\di{p}\cdot{Q}\hspace{-1pt}+\hspace{-1pt}\di{p}^2\hspace{-1pt}/\hspace{-1pt}2)
+
\delta(\di{p}\cdot{Q}\hspace{-1pt}-\hspace{-1pt}\di{p}^2\hspace{-1pt}/\hspace{-1pt}2)
\big]
f(-\di{p}^2,p_\u\hspace{-2pt}\cdot\hspace{-1pt}Q,p_\s\hspace{-2pt}\cdot\hspace{-1pt}Q),
\eeq
where $Q$ is a physical $\i$ particle momentum with energy $Q^0=E\hspace{-1pt}(\3{Q})=({\3{Q}^2+m_\i^2})^{1/2}$. The negative-energy solution of the $Q^2$ delta-function is outside the physical phase-space if the $\s$ particle is stable, as previously explained.

The spacetime structure of the collision process is encoded in the Fourier transform of the collision function. Therefore it is convenient to employ the Fourier representation of the delta-function
\beq
\delta(\di{p}\cdot{Q}\hspace{-1pt}+\hspace{-1pt}\di{p}^2\hspace{-1pt}/\hspace{-1pt}2)
+
\delta(\di{p}\cdot{Q}\hspace{-1pt}-\hspace{-1pt}\di{p}^2\hspace{-1pt}/\hspace{-1pt}2)=\frac{1}{\pi m_\i}
\BIn{4}{-\infty}{+\infty}{d\tau}e^{-i\tau[\di{p}\cdot{Q}]/m_\i}\cos\left(\frac{\tau\di{p}^2}{2m_\i}\right)
\,.
\eeq
This step introduces a parameter $\tau$ that represents the proper time of the $\i$ particle trajectory similarly to what done in eq.\,(\ref{eq:distour}) for the leading-order calculation. At variance with eq.\,(\ref{eq:distour}), $\tau$ is integrated on the whole real line and not only on the negative axis. This is in contrast with the semi-classical interpretation because the decay must happen before the scattering. However, changing sign to $\tau$ is equivalent to $\di{p}\to-\di{p}$ and the $Q$ integral stays invariant under this transformation.\,\footnote{As explained in the previous section, there exists a Lorentz transformation that changes the sign of $\di{p}$ while leaving the classical momenta invariant. Changing variable in the $Q$ integral by this transformation effectively induces a sign change of $\di{p}$ in the integrand.} Therefore the integral on $\tau>0$ is equal to the integral for negative $\tau$ and the collision function can be written as 
\beq\label{eq:sana3}
{\mathfrak{C}}_{\slashed{a}}(S,-\di{p}^2)\hspace{-1pt}=\hspace{-1pt}\frac{1}{2 m_\i}\,
\BIn{4}{-\infty}{0}d\tau
\,
\In{5}\frac{d^3\hspace{-1pt}\3{Q}}{(2\pi)^32E(\3{Q})}\,
e^{-i\tau[\di{p}\cdot{Q}]/m_\i}\cos\left(\frac{\tau\di{p}^2}{2m_\i}\right)
f(-\di{p}^2,p_\u\hspace{-2pt}\cdot\hspace{-1pt}Q,p_\s\hspace{-2pt}\cdot\hspace{-1pt}Q).
\eeq

The final step is to insert ${\mathfrak{C}}_{\slashed{a}}$ into eq.\,(\ref{eq:MAS3}) to obtain the transition probability. This is the exact non-analytic contribution to the transition probability that we can now expand semi-classically around $\di{p}=0$. At the leading order, the only dependence of ${\mathfrak{C}}_{\slashed{a}}$ on $\di{p}$ is in the exponential. It combines with the $\di{p}_\u\hspace{-1pt}\cdot\hspace{-1pt}{x}$ exponential of eq.\,(\ref{eq:MAS3}) and gives us the Wigner density of the $\u$ beam evaluated at $x+\tau Q/m_\i$. The Wigner density of $\s$ is evaluated at $x$ and we recover the structure of eq.\,(\ref{eq:ldsp})
\beq
{\mathcal{P}}_\s(x,\3{P}_\s)\times
{\mathcal{P}}_\u(x\hspace{-1pt}+\hspace{-1pt}\tau Q/m_\i,\3{P}_\u)\,.
\eeq
All the other factors of eq.\,(\ref{eq:ldsp}) are also recovered using the zeroth order expansion in in $\di{p}$ of the function $f$ that gives 
\beq
f(0,P_\u\hspace{-2pt}\cdot\hspace{-1pt}Q,P_\s\hspace{-2pt}\cdot\hspace{-1pt}Q)=16 (2\pi)^3 m_\u E(\3{q})\frac{d^3\Gamma_\D\hspace{-1pt}}{d^3\3{q}}\left[{\scriptstyle{\sqrt{(Q\cdot P_\s)^2-m_\i^2m_\s^2}}}\;\sigma_\P\right]\,.
\eeq

The terms of higher order in the classical expansion emerge from the Taylor series expansion of $f$ and of the cosine factor in eq.\,(\ref{eq:sana3}). Expanding of $f$ brings powers of $-\di{p}^2$ that correspond to powers of the $\Box$ operator acting on one of the two Wigner densities. The spacetime structure of these corrections is
\beq
{\mathcal{P}}_\s(x,\3{P}_\s)\times\Box_x^n{\mathcal{P}}_\u(x\hspace{-1pt}+\hspace{-1pt}\tau Q/m_\i,\3{P}_\u)\,.
\eeq
The expansion of the cosine brings even powers of $\tau\,\di{p}^{2}$ and corrections proportional to 
\beq
{\mathcal{P}}_\s(x,\3{P}_\s)\times
\bigg(\frac{\tau}{m_\i}\bigg)^{2n}\,\Box_x^{2n}{\mathcal{P}}_\u(x+\tau Q/m_\i,\3{P}_\u)\,.
\eeq

These results show the connection between non-analyticities in the collision function and long-distance effects in the transition probability that can be interpreted as the result of the decay of the $\u$ particle producing $\i$, the propagation of $\i$ over a macroscopic distance and eventually the scattering on $\s$. The correspondence is not limited to the leading order in the classical expansion but it extends to the quantum corrections. On the contrary, we saw in section\,\ref{sec:sdt} that analytic terms produce effects that are local in $x$ at all orders in the semi-classical expansion and correspond to the direct interaction between $\u$ and $\s$. This marks a sharp physical distinction between analytic and non-analytic terms in the collision function.

\subsection{Explicit calculation}
\label{sec:exc}

If $\D$ and $\P$ are single-particle states and the amplitudes ${\mathcal{M}}_{\D,\P}$ are constants (which we set to 1) the collision function of the $t$-channel process in figure\,\ref{diag1} is 
\beq\label{eq:sf2t2}
{\mathfrak{C}}_2 \hspace{-2pt}= \In{4} \d{\rm{Ph}}_{2}(p_{\rm{tot}})\, \Pi(\di{t},t)\hspace{-2pt}
= \In{4} \d{\rm{Ph}}_{2}(p_{\rm{tot}}) \frac{1}{2(\di{t} + i \epsilon)} \hspace{-1pt} \left[\frac{1}{t \hspace{-1pt}-\hspace{-1pt} m_\i^2\hspace{-1pt} -\hspace{-1pt} \di{t} \hspace{-1pt}- \hspace{-1pt} i \epsilon} \hspace{-1pt}-\hspace{-1pt} \frac{1}{t \hspace{-1pt}-\hspace{-1pt} m_\i^2 + \di{t} \hspace{-1pt}+\hspace{-1pt} i \epsilon}\right]\hspace{-1pt}.
\eeq
This can be computed analytically and the result can be also verified by numerical integration for finite but small $\epsilon$ regulator. The explicit calculation will offer a cross-check of the general formulas worked out in the previous sections, but it will also reveal new structures.

The 2-body kinematics is best visualized in the centre of mass frame, where the total mean momentum $p_{\rm{tot}}=p_{\u}+p_{\s}$ is at rest and the initial state mean 3-momenta $\3{p}_{\u}$ and $\3{p}_{\s}$ are taken parallel to the $z$-axis. In this frame $\di{p}$ has vanishing time and $z$ components (since it is orthogonal to ${p}_{\u,\s}$) and it can be aligned with the $x$-axis by a rotation around $z$. The norm $|\3{k}_{\hspace{-1pt}*}|$ of the final particles 3-momenta is determined in terms of $s$ and of the masses $\mu_\D$ and $\mu_\P$ by the familiar final-state K\"all\'en function $\Lambda_F$. The norm $|\3{p}_{\hspace{-1pt}*}|$ of the initial state 3-momenta $\3{p}_{\u,\s}$ is instead not determined by the initial-state $\Lambda_I$ but it also depends on $\di{p}^2$ because $p_{\u,\s}^2=m_{\u,\s}^2-\di{p}^2/4$. The definition of the K\"all\'en functions and the expressions for $|\3{k}_{\hspace{-1pt}*}|$ and $|\3{p}_{\hspace{-1pt}*}|$ are reported in table\,\ref{tab:2bpp}. Notice that $|\3{p}_{\hspace{-1pt}*}|^2$ is automatically positive since $s$ is bounded by eq.\,(\ref{eq:bons}). Positivity of $|\3{k}_{\hspace{-1pt}*}|^2$ requires $s$ to be above the threshold for the production of the final state particles
\beq
s>(\mu_\P+\mu_\D)^2.
\eeq

\begin{table}[t]
    \centering
    \fbox{%
  \begin{minipage}{0.9\textwidth}
  \vspace{-.5em}
\begin{eqnarray}
&\Lambda(\alpha, \beta, \gamma) \equiv \alpha^2 + \beta^2 + \gamma^2 - 2 \alpha \beta - 2 \alpha \gamma - 2 \beta \gamma = (\alpha-\beta-\gamma)^2-4\beta\gamma\,,\nonumber\\[4pt]
&\Lambda_F\equiv\Lambda(s,\mu_\D^2,\mu_\P^2)\,,\qquad\qquad
\Lambda_I\equiv\Lambda(s,m_\u^2,m_\s^2)
\nonumber\\[4pt]
&\displaystyle
|\3{k}_{\hspace{-1pt}*}|^2=\frac{\Lambda_F}{4s}\,,\qquad\;
|\3{p}_{\hspace{-1pt}*}|^2=
\frac{\Lambda(s,p_\u^2,p_\s^2)}{4s}=
\frac{\Lambda_I}
{4s}+\frac{\di{p}^2}{4}=\frac{p_\u^2p_\s^2}{m_\u^2m_\s^2}\frac{\Lambda(S,m_\u^2,m_\s^2)}{4s}\,.
\nonumber
\end{eqnarray}
\vspace{-.5em}
  \end{minipage}%
}
    \caption{The definition of the K\"all\'en functions and their relation with the momenta in the centre of mass frame. The last equality is derived from eq.\,(\ref{eq:ClS}).  
    \label{tab:2bpp}}
\end{table}

The two free variables that describe the final state momenta can be chosen to be the manifestly Lorentz-invariant variables $t$ and $\di{t}$. In the centre of mass frame $t$ determines---see e.g.\,eq.\,(\ref{eq:ttt})---the $z$-component of the momenta and $\di{t}$ determines the $x$-component. The $y$-component is also determined since $|\3{k}_{\hspace{-1pt}*}|$ is fixed, but only up to a sign ambiguity. The two halves of the phase-space with positive or negative $y$-component have to be treated separately and the variables $t$ and $\di{t}$ are valid coordinates in each of the two. The allowed ranges of $t$ and $\di{t}$ can be obtained by requesting that the squared sum of the $x$- and $z$-components of the momenta do not exceed $|\3{k}_{\hspace{-1pt}*}|^2$
\begin{eqnarray}
\label{eq:bottt}
&\displaystyle\di{t}^2 \leq - \frac{\Lambda_F}{4s}\di{p}^2\equiv(\di{t}_+)^2\,,\qquad\quad
t\in[t_-(\di{t}),t_+(\di{t})]\,,&\\
&\displaystyle t_\pm\equiv a\pm b= \frac{m_\s^2\hspace{-1pt} + m_\u^2\hspace{-1pt} + \mu_\D^2 + \mu_\P^2\hspace{-1pt} - s}{2} + \frac{(m_\s^2 \hspace{-1pt} - m_\u^2)(\mu_\D^2 \hspace{-1pt}- \mu_P^2)}{2s} \pm 
\sqrt{
\frac{\Lambda_I}
{s}\hspace{-1pt}+{\di{p}^2}
}
\sqrt{\frac{\Lambda_F}{4 s} \hspace{-1pt} + \frac{\di{t}^2}{\di{p}^2}}.&
\nonumber
\end{eqnarray}
Notice that $b$ is real and positive-definite, while $a$ can assume both signs depending on $s$ and on the masses. 

The integrand in eq.\,(\ref{eq:sf2t2}) features singularities on the two lines $t=m_\i^2\pm\di{t}$, which corresponds to the configurations where either $\0{q}$ or $\1{q}$ is on-shell. The lines belong to the accessible phase-space provided $t=m_\i^2\pm\di{t}$ belongs to the allowed range for $t$, namely if
\beq
 \left|a - m_\i^2 \pm \di{t}\right| \leq {b}\,,
\eeq
for some allowed $\di{t}$ in the range specified in eq.\,(\ref{eq:bottt}). The condition is equivalent to
\beq\label{eq:alpha}
\alpha \equiv - \frac{4s}{\Lambda_F}(a - m_\i^2)^2 + \frac{\Lambda_I}{s} \geq 0\,.
\eeq
Notice that $\alpha$ depends only on $s$ and not on $\di{p}^2$. 

A stronger condition is required for $\0{q}$ and $\1{q}$ to go on-shell at the same point of the phase-space. This is the configuration $t=m_\i^2$ and $\di{t}=0$ and it belongs to the phase-space only provided
\beq
\alpha \geq - \di{p}^2\,.
\eeq

The standard phase-space volume element formula that employs spherical coordinates is easily expressed in the variables $t$ and $\di{t}$ by a change of coordinates. The result is 
\beq\label{phsttt}
\d{\rm{Ph}}_{2}(p_{\rm{tot}}) = \frac{1}{8 \pi^2} \frac{1}{\sqrt{s}} \frac{1}{\sqrt{- \di{p}^2}} \frac{dt\, d \di{t}}{\sqrt{(t - t_-)(t_+ - t)}},
\eeq
where an additional factor $2$ is included to take into account that $t$ and $\di{t}$ cover only half of the phase space as previously explained.

We now evaluate eq.\,(\ref{eq:sf2t2}), starting from the integral over $t$ that gives
\beq
\BIn{10}{t_-}{t_+}\hspace{1em}dt\, \frac{1}{\sqrt{(t - t_-)(t_+ - t)}} \frac{1}{t - m_\i^2 \mp \di{t} \mp i \epsilon}
= \pm \frac{\pi i}{\sqrt{b^2 - (a - m_\i^2 \mp \di{t} \mp i \epsilon)^2}}\,.
\eeq
This is obtained by the standard technique of choosing the square-root branch with a cut along the segment $[t_-,t_+]$ and considering a contour that encloses the cut. The integral is then determined by the residue at the simple pole at $t=m_\i^2\pm\di{t}\pm i\epsilon$. Substituting this result into eq.\,(\ref{eq:sf2t2}) one finds
\beq
{\mathfrak{C}}_2 = \frac{i}{16 \pi} \frac{1}{\sqrt{s}} \frac{1}{\sqrt{-\di{p}^2}} \BIn{4}{-\di{t}_+}{+\di{t}_+} \hspace{0.5em} \frac{d \di{t}}{\di{t} + i \epsilon} \frac{1}{\sqrt{b^2 - (a - m_\i^2 - \di{t} - i \epsilon)^2}} + \textrm{h.c.}\,.
\eeq

The evaluation of the $\di{t}$ integral is not straightforward. It is easy to find a formal expression for the primitive, but this involves logarithms and square roots with branch cuts that cross the integration domain and the corresponding discontinuities need to be taken into account. The final result can be cast in the compact form 
\beq\label{eq:sf0}
{\mathfrak{C}}_2=
\displaystyle\frac{1}{2 \pi} \frac{1}{\sqrt{\Lambda_F}} \frac{1}{\sqrt{-\di{p}^2}} \frac{1}{\sqrt{-\alpha - \di{p}^2}}\,\textrm{arcsinh}\hspace{-2pt}\left(\sqrt{-\Theta(\alpha) - {\di{p}^2}/{|\alpha|}}\right)\,,
\eeq
where the parameter $\alpha=\alpha(s)$ is defined in eq.\,(\ref{eq:alpha}). 
A more useful equivalent expression for ${\mathfrak{C}}_2$ can be obtained by splitting the $(\alpha,-\di{p}^2)$ plane---or equivalently $(s,-\di{p}^2)$---into the regions where the $t$-channel momenta can go on-shell simultaneously, the one where they can go on-shell separately and the one where they are always off-shell. The collision function takes a different analytic form in each region:
\beq\label{eq:sf_explicit}
{\mathfrak{C}}_2 = 
\left\{
\begin{array}{l@{\qquad \textrm{for} \quad}l}
   \displaystyle\frac{1}{2 \pi} \frac{1}{\sqrt{\Lambda_F}} \frac{1}{\sqrt{-\di{p}^2}} \frac{1}{\sqrt{\di{p}^2 + \alpha}}\left[\frac{\pi}{2}\hspace{-1pt} -\hspace{-1pt} \arcsin\hspace{-2pt}\left(\sqrt{-{\di{p}^2}/{\alpha}}\right)\right] & \alpha > - \di{p}^2,\\
   \displaystyle\rule{0pt}{2.25em}\frac{1}{2 \pi} \frac{1}{\sqrt{\Lambda_F}} \frac{1}{\sqrt{-\di{p}^2}} \frac{1}{\sqrt{-\alpha - \di{p}^2}}\,\textrm{arcsinh}\hspace{-2pt}\left(\sqrt{-1 -\di{p}^2/\alpha}\right) & 0 < \alpha < -\di{p}^2,\\
   \displaystyle\rule{0pt}{2.25em}\frac{1}{2 \pi} \frac{1}{\sqrt{\Lambda_F}} \frac{1}{\sqrt{-\di{p}^2}} \frac{1}{\sqrt{-\alpha - \di{p}^2}}\,\textrm{arcsinh}\hspace{-2pt}\left(\sqrt{{\di{p}^2}/{\alpha}}\right) & \alpha < 0.
\end{array}
\right.
\eeq

\begin{figure}[t]
\centering
\includegraphics[width=.50\textwidth]{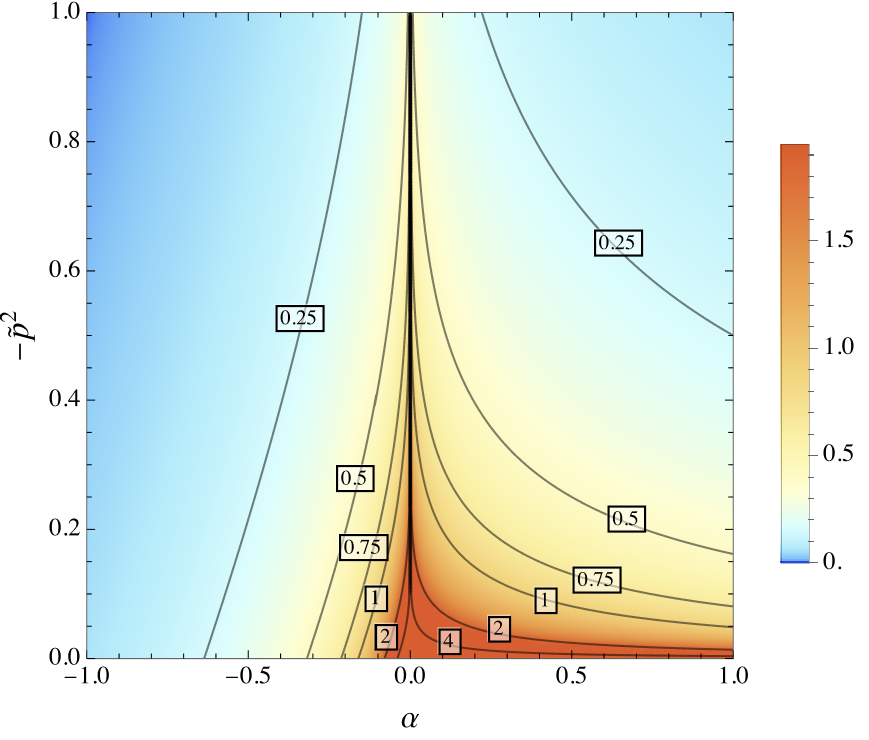}
\hfill
\raisebox{2.em}{\includegraphics[width=.48\textwidth]{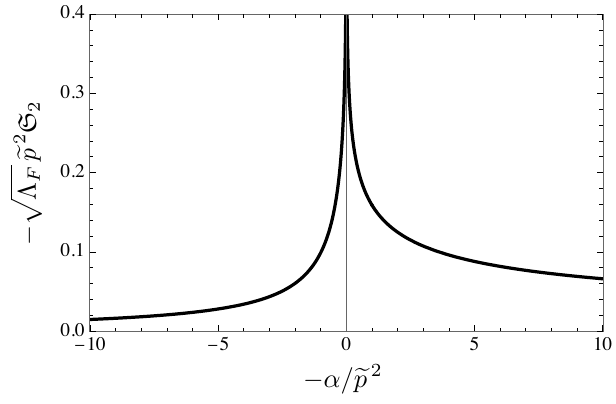}}
\caption{{\bf{Left:}} Plot of $\sqrt{\Lambda_F}\, {\mathfrak{C}}_2$ in the $(\alpha, -\widetilde{p}^2)$ plane. {\bf{Right:}} Plot of $-\sqrt{\Lambda_F}\, \widetilde{p}^2 {\mathfrak{C}}_2$ as a function of $-\alpha/\widetilde{p}^2$.}\label{fig:S2}
\end{figure}

This explicit formula matches the general expectations on the analyticity of the collision function at $\di{p}=0$. If $\alpha(s)$ is set to be negative the on-shell configuration is kinematically forbidden and ${\mathfrak{C}}_2$ is analytic at $\di{p}\hspace{-1pt}=\hspace{-1pt}0$. The relevant expansion for the classical limit is the one at fixed classical $S$ and not at fixed $s$, however $s(S,\di{p}^2)$ is an analytic function of $\di{p}^2$. Therefore ${\mathfrak{C}}_2$ is analytic at $\di{p}\hspace{-2pt}=\hspace{-2pt}0$ also at fixed $S$, provided $\alpha(S)\hspace{-3pt}<\hspace{-3pt}0$. If instead $\alpha(s)$ (or $\alpha(S)$) is positive the behaviour of ${\mathfrak{C}}_2$ for sufficiently small $\di{p}$ is given by the first line of eq.\,(\ref{eq:sf_explicit}). The $\pi/2$ term in the square bracket produces a term that is not analytic at $\di{p}\hspace{-1pt}=\hspace{-1pt}0$
\beq
 {\mathfrak{C}}_{2,\slashed{a}} = \frac{1}{4} \frac{1}{\sqrt{\Lambda_F}} \frac{1}{\sqrt{-\di{p}^2}} \frac{1}{\sqrt{\di{p}^2 + \alpha}}\,,
\eeq
while the contribution associated with the $\arcsin$ is analytic. We have verified that the non-analytic contribution is exactly reproduced by integrating the $\Pi_{\slashed{a}}$ propagator defined in eq.~(\ref{eq:Pi_na}) consistently with the findings of the previous sections.

In the previous sections we also derived in eq.\,(\ref{eq:sigmaan}) the general expression of the analytic contribution to the collision function, $\mathfrak{C}_a$, after subtracting the non-analytic terms. The zeroth-order expansion of $\mathfrak{C}_a$ is particularly interesting because it is associated with the unstable particle scattering cross section as in eq.\,(\ref{eq:scxsIP}). In the case under consideration of $2\to2$ scattering with constant $\D$ and $\P$ amplitudes the second line of eq.\,(\ref{eq:sigmaan}) vanishes and $\mathfrak{C}_a$ uniquely emerges from the $\Pi_{\rm{r}}$ part of the squared propagator. We claimed that this contribution can be obtained from the regular scattering theory calculation supplemented with the Hadamard finite-part regularization\,(\ref{eq:simppr}) for the $t$-channel singularity. The calculation is performed in the familiar phase-space where $\di{p}$ is effectively zero and the Mandelstam $t$ reads
\beq
t = a^0 + 2 |{\bf k}_*^0| |{\bf p}_*^0| \cos \theta\,,
\eeq
where $\theta$ is the polar angle in the centre of mass frame. The parameters $a$, $|{\bf k}_*|$ and $|{\bf p}_*|$ are defined as before but setting $\di{p}=0$ and consequently $s=S$. We obtain
\beq
{\mathfrak{C}}_{2,a}(S,0)\hspace{-1pt} =\hspace{-2pt} \frac{1}{8 \pi} \frac{|{\bf k}_*^0|}{\sqrt{S}} \In{4} d \hspace{-1pt}\cos\hspace{-1pt}\theta\, {\textrm{Pf}}\frac{1}{(a^0 \hspace{-1pt}- m_\i^2 \hspace{-2pt}+ \hspace{-1pt} 2 |{\bf k}_*^0| |{\bf p}_*^0| \cos \theta)^2}\hspace{-2pt}=\hspace{-2pt}
\frac{1}{8 \pi} \frac{|{\bf k}_*^0|}{\sqrt{S}} \frac{2}{(a^0\hspace{-2pt} - \hspace{-1pt} m_\i^2)^2\hspace{-3pt} -\hspace{-1pt} 4 |{\bf k}_*^0|^2 |{\bf p}_*^0|^2}
,\nonumber
\eeq
where the evaluation of the finite part is trivial using eq.\,(\ref{eq:pf2}) with constant $\phi$. This result is consistently equal to the zeroth order expansion of eq.\,(\ref{eq:sf_explicit})
\beq\label{eq:s2a0}
{\mathfrak{C}}_{2,a}(S,0)
 = - \frac{1}{2 \pi} \frac{1}{\sqrt{\Lambda_F^0}} \frac{1}{\alpha(S)}\,.
\eeq
Clearly for $\alpha<0$ the singularity plays no role, the standard calculation is well-defined and by plugging eq.\,(\ref{eq:s2a0}) into eq.\,(\ref{eq:scxsIP}) we recover the standard cross section. 

Eq.\,(\ref{eq:s2a0}) is negative when $\alpha>0$, producing a negative cross section. This does not result in a negative transition probability because the collision function for $\alpha>0$ is dominated by the non-analytic term, in the classical limit, and the transition probability is correspondingly dominated by the (positive) long-distance term. The cross section provides the leading-order short-distance contribution to the transition probability, which however is only a subleading correction to the total probability. The total probability~(\ref{eq:probability}) is positive by definition, therefore this negative correction overcoming the positive leading order merely signals a breakdown of the classical expansion. The physical effect of the negative cross section is to lower the probability in comparison with the classical long-distance prediction. The effect is associated with the short-distance direct interaction of the unstable particle with the other beam, yet it is a genuine quantum correction that cannot be described in classical probabilistic terms because the cross section is negative. 

Besides confirming the results obtained in the previous sections, eq.\,(\ref{eq:sf_explicit}) also reveals new aspects. In particular it shows that the radius of convergence of the ${\mathfrak{C}}_2$ expansion around $\di{p}^2=0$ is proportional to $|\alpha|$. This is physically interesting because the $\di{p}^2$ corrections are new quantum effects in the transition probability that cannot be described in classical probabilistic terms as the scattering (for $\alpha<0$) or the decay plus scattering (for $\alpha>0$) of point-like classical particles. The fact that they are enhanced as $\di{p}^2/|\alpha|$ at small $\alpha$ could facilitate their detection in experiments. Notice that we could have anticipated, based on appendix\,\ref{sec:dist}, the emergence of enhanced corrections close to the threshold (i.e., $\alpha=0$) at which the $t$-channel particle can go on-shell. If $\alpha$ is small, the singularity approaches the boundary of the phase-space and the radius of convergence of the distribution expansion is proportional to the distance between the singularity and the boundary.

The explicit result also allows us to investigate the properties of the collision function in the entire $(\alpha,-\di{p}^2)$ plane. The function is represented in figure~\ref{fig:S2}. It is analytic everywhere except the line $\alpha=0$ and the half-line $-\di{p}^2=0$ for positive $\alpha$, where it is singular. In particular it is analytic at $\alpha=-{\di{p}}^2$, which is easily established using eq.\,(\ref{eq:sf0}) for $\alpha>0$. The singularities of the collision function do not correspond to singularities in the physically observable transition probability. The transition probability is the integral of the collision function over $\di{p}$, but also over the spread in momentum of the colliding beams. The latter is effectively an integral over $\alpha$. It is easy to see that the singularities of the collision functions are integrable such that the transition probability is finite. As discussed in the previous sections, the singularity at $\di{p}^2=0$ and $\alpha\neq0$ is integrable in $\di{p}$ because the $\di{p}$ integration measure~(\ref{eq:vintpt}) is two-dimensional. The $\alpha=0$ singularity is logarithmic for $\di{p}\neq0$
\beq
 {\mathfrak{C}}_2 \simeq - \frac{1}{4 \pi} \frac{1}{\sqrt{\Lambda_F}} \frac{1}{\di{p}^2} \log \left(-4 \frac{\di{p}^2}{|\alpha|}\right) + {\cal O}(\alpha \log |\alpha/\di{p}^2|)\,,
\eeq
and therefore it has a finite $\alpha$ integral. Finally, the double integral in $\alpha$ and $\di{p}$ around the point $\alpha=0$ and $-\di{p}^2=0$ is finite.

The following global picture emerges from these results. If the classical momentum distribution of the colliding beams is localized well below the on-shell threshold, i.e.~at $\alpha(S)<0$, the transition probability is of the short-distance form and it matches the classical results up to corrections that are proportional (recall section~\ref{ssecCL}) to the inverse of the beam spatial extension squared divided by $|\alpha|$. Similarly, the classical long-distance transition probability due to the $\u$ particle decay is recovered up to small corrections if $\alpha(S)$ is localized well above the threshold. Instead, in a hypothetical experiment where $\alpha(S)$ is localized around zero with a small width large quantum corrections would emerge from the interplay between the on-shell and the off-shell regions. These effects are perfectly calculable in our framework and worth further studies.

\section{Conclusions}\label{sec:conc}

We have studied unstable particle collisions in QFT by a more complete approach to collision theory where the spacetime structure of the process emerges from the calculation rather than from a prior assumption like in the standard treatment. We obtained a complete characterization of the transition probability and we analysed its general structure at all orders in the classical beam expansion. In particular this enabled us to identify a term that contributes to the transition probability in the same way as the cross section in the standard classical formula. This provides the definition and the corresponding rules for computing the cross section for a process initiated by an unstable particle, which is ill-defined in the standard approach.

The spacetime collision theory formalism was described in section~\ref{sec:2}. An important improvement of previous literature was to introduce the notion of classical momentum, which led to the definition of a Lorentz-scalar Wigner density and a precise notion of the classical beam expansion. The final result in eq.\,(\ref{eq:MAS3}) for the transition probability involves the collision function ${\mathfrak{C}}(S,-\di{p}^2)$ and the classical expansion is the expansion of ${\mathfrak{C}}$ at fixed $S$ around $\di{p}=0$. If ${\mathfrak{C}}$ is analytic, the leading order reproduces the standard formula for the collision of classical point-like particles that undergo short-distance interactions. Higher order terms are also short-distance, but their contribution to the transition probability involves derivatives of the particle density. These contributions are quantum effects that do not possess a classical probabilistic interpretation.

Unstable particle collision was studied in section~\ref{sec:3}. In section~\ref{sec:sdt} we analysed the collision function and we isolated in eq.\,(\ref{eq:sigmaan}) the analytic term, disentangling it from the non-analyticities due to the $t$-channel singularity. This provided us with the definition of the unstable particle cross section in eq.\,(\ref{eq:scxsIP}). We saw in section~\ref{sec:exc} that this cross section can be negative, in which case its physical effect is a reduction of the transition probability in comparison with the classical long-distance prediction. It is a genuine quantum effect that cannot be interpreted in classical probabilistic terms. 

In section~\ref{sec:ldt} we studied the contribution to the transition probability from the non-analyticities at all orders in the classical expansion. We found that they all feature a long-distance structure that corresponds to the decay of $\u$ followed by the propagation and eventually the scattering of $\i$. However, only the leading term has a classical probabilistic interpretation. The others involve derivatives of the Wigner densities. Our results are for a general process featuring the $t$-channel singularity, however in section~\ref{sec:exc} we specialized to $2\to2$ scattering for an explicit cross-check.

Throughout this work we treated the unstable particle $\u$ as an incoming asymptotic state, which is inconsistent. The correct physical picture is that $\u$ is produced at some potentially large but finite time $-T$, it propagates until it crosses the other colliding beam and finally collides. The $\u$ particle production process should thus be taken into account in the calculation as already suggested in the literature~\cite{Ginzburg:1995js,Dams:2002uy}.

The lack of the production modelling shows up in our results for the transition probability at the leading order in the classical expansion, both in the short-distance~(\ref{eq:MASCL2p}) and in the long-distance~(\ref{eq:ldsp}) configurations. These expressions are identical in form to the result of a classical calculation, but they do not coincide exactly with the classical formula because ${\mathcal{P}}_\u$ evolves in time as the probability density of a freely-propagating beam of classical and stable particles. The evolution of the unstable $\u$ particle beam should instead take into account the attenuation due to decay and contain a survival probability factor $\exp[-\Gamma_\u m_\u T/P_\u^0]$, where $\Gamma_\u$ is the decay width and $T$ the time elapsed from the production. Our results are correct only if the width is small $\Gamma_\u\ll P_\u^0/(m_\u T)$. 

The $\u$ particle production process could be included straightforwardly in the spacetime collision theory calculation and the complete experimental setup described rigorously as a collision with 3 or more stable particles---those that produce $\u$, plus $\s$---in the initial state. Preliminary studies and similar calculations (see e.g.~\cite{Dobrev:2025crc}) show that it is relatively easy to extract the leading-order result recovering the survival probability. Studying the parametric scaling of the corrections is instead rather involved due to the interplay between the width, the separation between production and scattering regions, and the parameters that control the classical expansion. The inclusion of the finite width effects is thus left to future work.

A possible phenomenological application of our work is the calculation of scattering cross sections in experiments with muon beams such as $\mu^-\mu^+\to e^-\bar\nu_e W^+$ at a future high-energy muon collider. 
It would be interesting to compare with previous results~\cite{Melnikov:1996na,Melnikov:1996iu,Melnikov:1996ft} obtained with a cutoff in the phase space.

A last more speculative direction is to survey processes with muons or other unstable particles in the search for the design of an experimental setup where some of the novel quantum effects we identified could be actually observed. The simple estimate of section~\ref{ssecCL} shows that this is not an easy task: regular particle beams are deeply in the region of validity of the classical approximation. On the other hand, we saw in section\,\ref{sec:exc} that quantum effects are enhanced in the vicinity of the on-shell threshold potentially leading to a structured phenomenology that is worth exploring.

\section*{Acknowledgements}
This work was initiated during the GGI workshop ``Exploring the energy frontier with muon beams''.
We thank Kyu~Jung~Bae and Hyung~Do~Kim for collaboration at a preliminary stage of the project.  The work is 
part of the R\&D\&i project PID2023-146686NB-C31, funded by MICIU/AEI/10.13039/501100011033/ and by ERDF/EU. It is also supported by the Departament de Recerca i Universitats from Generalitat de Catalunya, Grup de Recerca ‘Grup de Física Teòrica UAB/IFAE’ (2025 SGR 01317). Z.L. is supported by the Department of Energy under Grant No.~DE-SC0011842 and in
part by a Sloan Research Fellowship from the Alfred P.~Sloan Foundation at the University
of Minnesota.

\appendix

\section{Notation guide}\label{app:notation}

Table~\ref{tab:notation} collects the main symbols and their defining equations or passages. 

\begingroup
\small
\setlength{\tabcolsep}{4pt}
\renewcommand{\arraystretch}{1.18}
\begin{longtable}{@{}>{\raggedright\arraybackslash}p{.22\textwidth}>{\raggedright\arraybackslash}p{.53\textwidth}>{\raggedright\arraybackslash}p{.19\textwidth}@{}}
\caption{Main symbols, meanings and definition locations.}\label{tab:notation}\\
\hline
\textbf{Symbol} & \textbf{Meaning} & \textbf{Definition}\\
\hline
\endfirsthead
\multicolumn{3}{@{}l}{\textit{Table~\ref{tab:notation}: continued}}\\
\hline
\textbf{Symbol} & \textbf{Meaning} & \textbf{Definition}\\
\hline
\endhead
\hline
\endfoot
\multicolumn{3}{@{}l}{\textbf{States, beams and momenta}}\\[3pt]
$\varphi$, $|\varphi\rangle$ & Wave-packet function and normalized one-particle state. & Eq.~(\ref{eq:state})\\
$w_\alpha$, $\rho$ & Probabilities of the beam's component states and their momentum-space density matrix; the beam need not be pure. & Eq.~(\ref{eq:DM})\\
$\0 p$, $\1 p$ & Positive-energy on-shell momenta of the ket and bra, respectively; circled momenta enter the amplitude, dotted momenta its conjugate. & Eqs.~(\ref{eq:state}), (\ref{eq:ltr})\\
$p$, $\di p$ & Mean momentum $(\0 p+\1 p)/2$ and momentum difference $\0 p-\1 p$; $p\cdot\di p=0$. & Eq.~(\ref{eq:meandifmom})\\
$P$, $E(\3 P)$ & Classical on-shell momentum $P=mp/\sqrt{p^2}$ and its energy $P^0=\sqrt{m^2+\3 P^2}$. & Eq.~(\ref{eq:cmom1})\\
$\mathcal P_C(x,\3 P)$ & Classical probability density in position and momentum at time $x^0$. & Eq.~(\ref{eq:PCdef})\\
$\mathcal P(x,\3 P)$ & Lorentz-scalar Wigner density; real and normalized, but not generally positive. & Eq.~(\ref{eq:Wddef})\\
$\kappa$, $\mu$ & Momentum-difference scale of a nearly classical beam, and the mass-gap scale used in its expansion. & Eqs.~(\ref{eq:supp}), (\ref{eq:dmsmu})\\
$q$, $\mathcal F$ & In the current-measurement discussion: particle charge and current form factor. This scalar $q$ is distinct from the exchanged four-momentum below. & Eqs.~(\ref{eq:avgcurr}), (\ref{eq:FFq})\\[4pt]
\multicolumn{3}{@{}l}{\textbf{Collisions and subprocesses}}\\[3pt]
$\u$, $\s$, $\i$ & Incoming particles $\u,\s$ and the exchanged decay product $\i$; $\u$ is unstable in section~\ref{sec:3}. & Section~\ref{se:sds}; eqs.~(\ref{eq:decc}), (\ref{eq:prod})\\
$\D$, $\P$ & Single- or multi-particle final states in $\u\to\i\,\D$ and $\i\,\s\to\P$. & Eqs.~(\ref{eq:decc}), (\ref{eq:prod})\\
$m_{\u,\s,\i}$; $\mu_{\D,\P}$ & Particle masses; invariant masses of the final-state systems, $K_{\D,\P}^2=\mu_{\D,\P}^2$. & Figure~\ref{diag1}; text before eq.~(\ref{eq:tdefs})\\
$K_\D$, $K_\P$ & Total four-momenta of the two final-state systems, each a sum of its constituent $k$ momenta. & Text before eq.~(\ref{eq:tdefs})\\
$p$ (probability) & Integrated transition probability; distinguished by context from the mean four-momentum $p^\mu$. & Eq.~(\ref{eq:probability})\\
$\mathcal M$; $\mathcal M_\D,\mathcal M_\P$ & Scattering amplitude; amplitudes for the decay and scattering subprocesses. & Eqs.~(\ref{eq:Tm}), (\ref{eq:ampfact})\\
$\mathfrak{C}$ & collision function obtained by integrating the amplitude times its conjugate over the final-state phase space. & Eq.~(\ref{eq:sF})\\
$s$, $S$ & Squared total mean and classical momenta: $s=(p_\u+p_\s)^2$, $S=(P_\u+P_\s)^2$. The classical expansion holds $S$ fixed. & Eq.~(\ref{eq:meanS}); text and eq.~(\ref{eq:ClS})\\
$\sigma$, $\di v_{\rm rel}$ & Cross section and flux velocity factor. For unstable-particle scattering, $\sigma$ is defined from the analytic coefficient. & Eqs.~(\ref{eq:scxs}), (\ref{eq:MASCL2p}), (\ref{eq:scxsIP})\\
$t$, $\di t$ & Mean and half-difference of the exchanged virtualities: $(\0 q^2+\1 q^2)/2$ and $(\0 q^2-\1 q^2)/2$. Unlike $\di p$, $\di t$ includes a factor $1/2$. & Eq.~(\ref{eq:tdefs})\\
$\Pi$, $\Pi_{\rm r}$, $\Pi_{\rm i}$ & $\Pi$: product of the exchanged propagators in the amplitude and its conjugate. $\Pi_{\rm r}$, $\Pi_{\rm i}$: contributions generated by the real and imaginary terms in the partial-fraction brackets, respectively. & Eqs.~(\ref{eq:Pidef}), (\ref{eq:Pidecom}), (\ref{eq:Pidecomi})\\
$\mathfrak D$, $\mathfrak S$ & Subprocess amplitude--conjugate-amplitude products integrated over the $\D$ and $\P$ phase spaces; $\mathfrak S$ is not the Wigner density $\mathcal P$. & Eq.~(\ref{eq:PD})\\
${\mathfrak{C}}_a$, ${\mathfrak{C}}_{\slashed a}$ & Analytic and non-analytic components of the collision function in the stated separation. & Eqs.~(\ref{eq:sigmaan}), (\ref{eq:nan})\\
$\Pi_{\slashed a}$ & Double-delta kernel $\pi^2\delta(\di t)\delta(t-m_\i^2)$, generating $\mathfrak S_{\slashed a}$. & Eq.~(\ref{eq:Pi_na})\\
$Q$, $q$ (four-momenta) & Physical on-shell momentum of $\i$ in the propagation representation, and mean exchanged momentum $q=p_\u-K_\D$. & Eq.~(\ref{eq:Qdef}) and following text; eq.~(\ref{eq:qdef})\\
$\Gamma_\D$, $\sigma_\P$ & Partial decay width for $\u\to\i\,\D$ and cross section for $\i\,\s\to\P$. & Eqs.~(\ref{eq:partialwidth}), (\ref{eq:productionxs})\\
$\tau$ & Daughter-particle proper-time parameter along $x+\tau Q/m_\i$; $\tau<0$ places the decay before the collision at $x$. & Eq.~(\ref{eq:distour}) and following text\\[4pt]
\multicolumn{3}{@{}l}{\textbf{Scalar example, preparation and distributions}}\\[3pt]
${\mathfrak{C}}_2$ & Collision function for the $2\to2$ scalar example with constant subprocess amplitudes set to one. & Eq.~(\ref{eq:sf2t2})\\
$\Lambda$, $\Lambda_I$, $\Lambda_F$ & K\"all\'en function and its initial- and final-state forms. & Table~\ref{tab:2bpp}\\
$a$, $b$, $t_\pm$, $\di t_+$ & Midpoint and half-width of the allowed $t$ interval at fixed $\di t$, its endpoints $t_\pm=a\pm b$, and the maximal $|\di t|$. & Eq.~(\ref{eq:bottt})\\
$\alpha(s)$ & On-shell support parameter: single poles are accessible for $\alpha\geq0$, and the double cut for $\alpha\geq-\di p^2$. & Eq.~(\ref{eq:alpha}) and following equation\\
$T$, $\Gamma_\u$ & In the preparation discussion: elapsed coordinate time and total rest-frame width of the unstable parent; $T$ here is not the transition operator of eq.~(\ref{eq:Tm}). & Section~\ref{sec:conc}\\
$\eta$, $\phi(x)$ & Shift parameter and test function in the distribution expansion; $\phi$ is distinct from the wave-packet function $\varphi$. & Eq.~(\ref{eq:distr_def11})\\
$\epsilon$, ${\rm pv}$, ${\rm Pf}$ & Positive propagator regulator, Cauchy principal value and Hadamard finite part; the distribution identities use the $\epsilon\to0^+$ boundary value. & Eqs.~(\ref{eq:Pidef}), (\ref{eq:1/x^n_expansion}), (\ref{eq:pf2})\\
\end{longtable}
\endgroup

\section{Current and energy-momentum tensor measurements}\label{app:LM}

In section\,\ref{sec:2} we motivated our definition of the Wigner density and of the classical limit based on the measurement of the charge density on a single beam of particles. Here we extend the analysis to all the components of the charge current $J^\mu(x)$ and of the energy-momentum tensor $T^{\mu\nu}(x)$ operators. These quantities have the sharp physical interpretation of the density of charge ($J^0$), of charge flux ($J^i$), of four-momentum ($T^{0\mu}$) and of four-momentum flux ($T^{i\mu}$). Their classical expectation value is determined by the densities generated by a single free point-like particle of definite four-momentum $P$ and position $\vec{r}$
\beq
{\rm{j}}^\mu=\frac{q}{P^0}\,\delta^3(\3{x}-\vec{r})P^\mu\,,\qquad
{\rm{t}}^{\mu\nu}=\frac{1}{P^0}
\,\delta^3(\3{x}-\vec{r})P^\mu P^\nu\,,
\eeq
convoluted with the classical space-momentum probability density ${\mathcal{P}}_C\hspace{-1pt}(x,\3{P})$ that characterises the beam. This gives
\beq\label{eq:avgCL}
{\mathds{E}}_C[{J^\mu(x)}]=q\,
\In{4}d^3\hspace{-2pt}\3{P}\,\frac{P^\mu}{P^0}{\mathcal{P}}_C(x,\3{P}) 
\,,\qquad
{\mathds{E}}_C[{T^{\mu\nu}(x)}]=\In{4}d^3\hspace{-2pt}\3{P}\,\frac{P^\mu P^\nu}{P^0}{\mathcal{P}}_C(x,\3{P}) 
\,.
\eeq
These quantities are conserved as the space-momentum density obeys $P^\mu\partial_\mu {\mathcal{P}}_C=0$.

The exact expectation value is given by eq.\,(\ref{eq:avgloc}), where ${\mathcal{O}}(\0{\3{p}},\1{\3{p}})$ are the operator matrix elements between plane-wave states
\al{
\label{eq:me}
&\langle\1{p}|
J^\mu(0)
|\0{p}\rangle
=2{\cal{F}}(\di{p}^2)p^\mu,\\
&\langle\1{p}|
T^{\mu\nu}(0)
|\0{p}\rangle
=2{\cal{F}}_1(\di{p}^2)p^\mu p^\nu+{\cal{F}}_2(\di{p}^2)\big(
\di{p}^\mu\di{p}^\nu
-\di{p}^2\eta^{\mu\nu}
\big).\nonumber
}
Here we used Lorentz symmetry and current conservation to parametrize the matrix elements in terms of scalar form-factors. Matching the total charge and energy of the particle requires ${\cal{F}}(0)=q$ and ${\cal{F}}_1(0)=1$.

The classical limit is the one where the support of the density matrix is localized at $-\di{p}^2<\kappa^2$ with $\kappa$ small. We saw in section\,\ref{ssecCL} that in this kinematic regime all the components of $\di{p}$ are of order $\kappa$ and $p=P$ up to quadratic corrections. Furthermore, the form-factor can be expanded and eq.\,(\ref{eq:me}) becomes
\al{
&\langle\1{p}|
J^\mu(0)
|\0{p}\rangle
\simeq2\,qP^\mu+O(\kappa^2),\\
&\langle\1{p}|
T^{\mu\nu}(0)
|\0{p}\rangle
\simeq2P^\mu P^\nu+O(\kappa^2).\nonumber
}
In order to study the classical limit it is convenient to change variables in the integral~(\ref{eq:avgloc}), using the coordinates $\3{P}$ and $\di{\3{p}}$. Taking into account the Jacobian~(\ref{eq:intmeas}) this gives
\beq
{\mathds{E}}[{O(x)}]=
\In{4}d^3\hspace{-2pt}\3{P}\left[
\frac1{(2\pi)^6}\frac1{2P^0}
\In{3}d^3\hspace{-1pt}\di{\3{p}}\,
\frac{{\mathcal{O}}(\0{\3{p}},\1{\3{p}})}{2p^0}
\frac{p^2}{m^2}
\rho(\0{\3{p}},\1{\3{p}}) \,
e^{-i\di{p}\cdot{x}}
\right].
\eeq
Using the definition of the Wigner density in eq.\,(\ref{eq:Wddef}) we immediately recover the classical expectation up to quadratic corrections:
\beq
{\mathds{E}}[{J^\mu(x)}]\simeq q\,
\In{4}d^3\hspace{-2pt}\3{P}\,\frac{P^\mu}{P^0}{\mathcal{P}}(x,\3{P}) +O(\kappa^2)
\,,\qquad
{\mathds{E}}_C[{T^{\mu\nu}(x)}]\simeq 
\In{4}d^3\hspace{-2pt}\3{P}\,\frac{P^\mu P^\nu}{P^0}{\mathcal{P}}(x,\3{P}) +O(\kappa^2)
\,.\nonumber
\eeq

\section{The relativistic Wigner density}
\label{sec:relwd}

Our definition of the Wigner density differs from the one given in ref.~\cite{Kotkin:1992bj} and its main advantage\,\footnote{Another advantage is that it obeys the conservation equation in eq.\,(\ref{eq:WDP}) with on-shell $P$. Because of this its evolution in time matches the one of the classical space-momentum density.} is being a scalar function under the Lorentz group. In this appendix we compare the two definitions with a particular focus on their transformation properties.

\begin{table}[t]
    \centering
\begin{minipage}{0.45\textwidth}
\al{
&\3{P}=\sqrt{\frac{m^2}{p^2}}\frac{\0{\3{p}}+\1{\3{p}}}2\,,
\quad
\di{\3{p}}=\0{\3{p}}-\1{\3{p}}\,,
\nonumber\\
&\frac{p^2}{m^2}=\frac{(E(\0{\3{p}})+E(\1{\3{p}}))^2-
|\0{\3{p}}+\1{\3{p}}|^2
}{4 m^2}\,.\nonumber
}
\end{minipage}
\hfill
\begin{minipage}{0.5\textwidth}
\al{
&\0{\3{p}}=\sqrt{\frac{p^2}{m^2}}\3{P}+\di{\3{p}}/2\,,
\quad\1{\3{p}}=\sqrt{\frac{p^2}{m^2}}\3{P}-\di{\3{p}}/2\,,\nonumber\\
&\frac{p^2}{m^2}=1
+\frac{|\di{\3{p}}|^2}{4m^2}-\frac{(\3{P}\cdot\di{\3{p}})^2}{4m^2E^2\hspace{-1pt}(\3{P})}\,.\nonumber
}
\end{minipage}
    \caption{The transformation from circled/dotted to classical/tilded momenta and its inverse.  
    \label{tab:tras}}
\end{table}

We start by rewriting our definition~(\ref{eq:Wddef}) of the Wigner density as 
\beq\label{eq:wdd2}
{\mathcal{P}}(x,\3{P})=
\frac1{(2\pi)^6}\frac1{2{E\hspace{-1pt}(\3{P})}}
\In{3}d^3\hspace{-1pt}\di{\3{p}}\,
\frac{p^2}{m^2}
\,{\mathcal{R}}(\3{P},\di{\3{p}}) \,
e^{-i\di{p}\cdot{x}},
\eeq
where we have introduced 
\beq
{\mathcal{R}}(\3{P},\di{\3{p}})\equiv\rho\big(\0{\3{p}}(\3{P},\di{\3{p}}),\1{\3{p}}(\3{P},\di{\3{p}})\big)\,.
\eeq
The notation makes explicit the change of coordinates from $(\0{\3{p}},\1{\3{p}})$ to $(\3{P},\di{\3{p}})$---reported on the right panel of table\,\ref{tab:tras}---that needs to be operated on $\rho$ before the integration. The integral is performed over $\di{\3{p}}$ at fixed $\3{P}$. Of course also the temporal component of $\di{p}$, which appears in the exponent, must be expressed in the same coordinates using 
\beq\label{eq:pt0}
\di{p}^0\hspace{-2pt}(\3{P},\di{\3{p}})=\frac{\3{P}\cdot\di{\3{p}}}{E\hspace{-1pt}(\3{P})}\,.
\eeq
The same is true for the $p^2/m^2$ factor
\beq
\frac{p^2\hspace{-2pt}(\3{P},\di{\3{p}})}{m^2}=
1+\frac{\di{\3{p}}^2}{4m^2}-\frac{(\3{P}\cdot\di{\3{p}})^2}{4m^2E^2\hspace{-1pt}(\3{P})}\,.
\eeq
It should be noted that this factor is Lorentz-invariant and it approaches one in the classical limit. It could have been excluded---or included at a different power---from our definition without spoiling Lorentz invariance and the properties reported in eq.\,(\ref{eq:WDP}) or harming the correspondence with the classical space-momentum density in the classical limit.

Under a Lorentz transformation the circled and dotted 3-momenta transform as the space component of on-shell physical momenta with positive energy and mass $m$. Namely a change of frame on the six-dimensional space $(\0{\3{p}},\1{\3{p}})$ corresponds to the transformation
\beq
\0{\3{p}}\;\longrightarrow\;
\Lambda_m\hspace{-1pt}(\0{\3{p}})\,,\quad
\1{\3{p}}\;\longrightarrow\;
\Lambda_m\hspace{-1pt}(\1{\3{p}})\,.
\eeq
The density matrix is a Lorentz scalar, namely the Lorentz group acts as
\beq\label{eq:trex0}
\rho(\0{\3{p}},\1{\3{p}})
\;\longrightarrow\;
\rho(
\Lambda_m\hspace{-1pt}(\0{\3{p}})
,
\Lambda_m\hspace{-1pt}(\1{\3{p}})
)\,.
\eeq
The explicit form of $\Lambda_m$ in the case of a boost along the $z$ axis is
\beq\label{eq:trex}
\Lambda_m\hspace{-1pt}\big(\zu{\3{p}}\big)=\big\{
\zu{\3{p}}_\bot,
\gamma\,\zu{\3{p}}_z
-\beta\gamma\,E(\zu{\3{p}})
\big\}\,,\qquad
E(\zu{\3{p}}))=\sqrt{m^2+\zu{\3{p}}^2}\,,
\eeq
where $\beta$ is the boost velocity and $\gamma=1/\sqrt{1-\beta^2}$. This non-linear transformation acting on ${\zu{\3{p}}}$ induces a regular linear Lorentz transformation on the four-vectors ${\zu{p}}$. 

The effect of a Lorentz transformation on the new coordinates $\3{P}$ and $\di{\3{p}}$ can be worked out straightforwardly by acting with~(\ref{eq:trex0}) on the old coordinates in the left panel of table\,\ref{tab:tras} and using the right panel to switch back to the new coordinates. The result is
\al{
\3{P}\;&\longrightarrow\;
\big\{
{\3{P}}_\bot,
\gamma\,{\3{P}}_z
-\beta\gamma\,E({\3{P}})
\big\}=
\Lambda_m\hspace{-1pt}(\3{P})\,,\\
\di{\3{p}}
\;&\longrightarrow\;
\big\{
{\di{\3{p}}}_\bot,
\gamma\,{\di{\3{p}}}_z
-\beta\gamma\,\frac{\3{P}\cdot\di{\3{p}}}{E({\3{P}})}
\big\}\equiv\di{\Lambda}({\3{P}},\di{\3{p}})\,.\nonumber
}
The classical 3-momentum $\3{P}$ transforms---just like $\0{\3{p}}$ and $\1{\3{p}}$---as the space component of a four-momentum with mass $m$ and positive energy. This result emerges non-trivially from the explicit calculation but is automatic in the formalism because $\3{P}$ is indeed defined as the space component of the classical four-momentum $P^\mu$. The ${\di{\3{p}}}$ transformation is instead the one of the space component of the four-vector $\di{p}$, whose temporal component is ${\3{P}\hspace{-2pt}\cdot\hspace{-2pt}\di{\3{p}}}/{E({\3{P}})}$  as in eq.\,(\ref{eq:pt0}). It is possible to check that
\beq
\di{p}(\Lambda_m\hspace{-1pt}(\3{P}),\di{\Lambda}({\3{P}},\di{\3{p}}))=
\big\{\di{p}^0\hspace{-1pt}(\Lambda_m\hspace{-1pt}(\3{P}),\di{\Lambda}({\3{P}},\di{\3{p}})),\di{\Lambda}({\3{P}},\di{\3{p}})\big\}
=
\Lambda\di{p}\,.
\eeq

The function ${\mathcal{R}}(\3{P},\di{\3{p}})$ that appears in the Wigner density definition~(\ref{eq:wdd2}) is a scalar and the other factors are invariant. The Jacobian of the change of variables $\di{\3{p}}\to \di{\3{p}}^\prime = \di{\Lambda}({\3{P}},\di{\3{p}})$ is readily found to be
\beq
d^3\hspace{-1pt}\di{\3{p}}^\prime=d^3\hspace{-1pt}[\di{\Lambda}({\3{P}},\di{\3{p}})]
=\frac{E(\Lambda_m\hspace{-1pt}(\3{P}))}{E(\3{P})}
d^3\hspace{-1pt}\di{\3{p}}\,,
\eeq
compatibly with eq.\,(\ref{eq:invm}). Therefore, under a Lorentz transformation 
\al{
{\mathcal{P}}(x,\3{P})
\longrightarrow\;
&\frac1{2{E\hspace{-1pt}(\3{P})}}
\In{3}
\frac{d^3\hspace{-1pt}\di{\3{p}}}{(2\pi)^6}
\,
\frac{p^2\hspace{-2pt}(\3{P},\di{\3{p}})}{m^2}
\,{\mathcal{R}}(\Lambda_m\hspace{-1pt}(\3{P}),
\di{\Lambda}({\3{P}},\di{\3{p}}))\,
e^{-i\big[\di{p}(\3{P},\,\di{\3{p}})\big]\cdot{x}}\\
=&\frac1{2{E\hspace{-1pt}(\Lambda_m\hspace{-1pt}(\3{P}))}}
\In{3}
\frac{d^3\hspace{-1pt}\di{\3{p}}^\prime}{(2\pi)^6}\,
\frac{p^2(\Lambda_m\hspace{-1pt}(\3{P}),\di{\3{p}}^\prime)}{m^2}
\,{\mathcal{R}}(\Lambda_m\hspace{-1pt}(\3{P}),\di{\3{p}}^\prime)\,
e^{-i\big[\di{p}(\Lambda_m\hspace{-1pt}(\3{P}),\,\di{\3{p}}^\prime)\big]\cdot{(\Lambda x)}}\nonumber\\
=&
{\mathcal{P}}(\Lambda x,\Lambda_m\hspace{-1pt}(\3{P})),
\nonumber
}
which shows explicitly that our definition of the Wigner density is a Lorentz scalar

We turn now to the standard definition~\cite{Kotkin:1992bj} of the Wigner density. After taking into account the different normalization of the plane-wave states and of the wave-packet function, it reads
\beq\label{eq:stdef}
{\mathcal{P}}_{\rm{st}}(x,\3{p})\equiv
\frac1{(2\pi)^6}
\In{3}d^3\hspace{-1pt}\di{\3{p}}\,
\frac1{2p^0}\left[
1-\left(
\frac{\di{p}^0}{2p^0}
\right)^2
\right]^{-1/2}
\hspace{-6pt}{\mathcal{R}}_{\rm{st}}(\3{p},\di{\3{p}}) \,
e^{-i\di{p}\cdot{x}},
\eeq
where we employed the relation $E(\0{\3{p}})E(\1{\3{p}})=(p^0)^2-(\di{p}^0)^2/4$ in order to express some energy factors and we defined
\beq
{\mathcal{R}}_{\rm{st}}(\3{p},\di{\3{p}})\equiv\rho\big(\0{\3{p}}(\3{p},\di{\3{p}}),\1{\3{p}}(\3{p},\di{\3{p}})\big).
\eeq
The integral is performed at fixed $\3{p}=(\0{\3{p}}+\1{\3{p}})/2$ and this variable (and not $\3{P}$) is the argument of the Wigner density. The relevant change of coordinates to be performed in $\rho$ before the integration is from the $(\0{\3{p}},\1{\3{p}})$ to the $(\3{p},\di{\3{p}})$ variables and the same variables must be used to express all the other terms in the integrand using
\al{\label{eq:p0pt0}
&2p^0\hspace{-2pt}(\3{p},\di{\3{p}})=E(\3{p}+\di{\3{p}}/2) + E(\3{p}-\di{\3{p}}/2)\,,\\
&
\di{p}^0\hspace{-2pt}(\3{p},\di{\3{p}})=
E(\3{p}+\di{\3{p}}/2) - E(\3{p}-\di{\3{p}}/2)\,.\nonumber
}

The main difference with our definition is that the integral is now performed on manifolds with constant $\3{p}$ and not with constant classical 3-momentum $\3{P}=m\3{p}/\sqrt{p^2}$. The problem is that in the $(\3{p},\di{\3{p}})$ coordinates the Lorentz group acts as 
\al{
\3{p}\;&\longrightarrow\;\frac{\Lambda_m\hspace{-1pt}({\3{p}}+\di{\3{p}}/2)+\Lambda_m\hspace{-1pt}({\3{p}}-\di{\3{p}}/2)}2\,,\\
\di{\3{p}}\;&\longrightarrow\;
\Lambda_m\hspace{-1pt}({\3{p}}+\di{\3{p}}/2)-\Lambda_m\hspace{-1pt}({\3{p}}-\di{\3{p}}/2)\,,\nonumber
}
and the transformation of $\3{p}$ depends also on $\di{\3{p}}$ and not only on $\3{p}$ itself. Therefore a manifold with constant $\3{p}=\3{p}_*$ is not mapped to another manifold $\3{p}=\3{p}_*^\prime$ that could provide the integration surface in the new frame. The standard definition~(\ref{eq:stdef}) is thus intrinsically not a scalar because it is based on a slicing of the six-dimensional space of momenta that is incompatible with Lorentz symmetry. Introducing the classical momentum solves the problem because the $\3{P}$ transformation does indeed depend solely on $\3{P}$.

\section{Distribution expansion}
\label{sec:dist}

In this appendix we consider a distribution depending on a parameter $\eta$ and the possibility of expanding it as a distributional series with a finite radius of convergence around $\eta=0$. Namely, of expanding in series the result of applying the distribution to a test function with suitable properties. We focus for definiteness on the distribution
\begin{equation}
\frac{1}{x - \eta + i \epsilon}\,,
\end{equation}
even if the results are of more general validity. We define the distribution acting as an integral over $x$ in the finite interval $[a,b]$, namely
\begin{equation}\label{eq:distr_def11}
    \left\langle \frac{1}{x - \eta + i \epsilon}, \phi(x) \right\rangle \equiv \int_{a}^{b} \frac{\phi(x)}{x - \eta + i \epsilon} dx\,.
\end{equation}
This corresponds to the integrals we deal with in the main text, which are performed on the phase-space that has finite volume. The extremes of integration as well as the test function $\phi$ can depend on $\eta$.

When $\epsilon$ is finite there is obviously no problem to expand using the geometric series
\begin{equation}
\frac{1}{x - \eta + i \epsilon} = \sum\limits_{k=0}^\infty \frac{\eta^k}{(x + i \epsilon)^{k+1}}\,.
\end{equation}
At fixed $x$, the series converges for
\begin{equation}
|\eta|<|x+i\epsilon|=\sqrt{x^2+\epsilon^2}.
\end{equation}
Therefore, if $|\eta|<\epsilon\leq \sqrt{x^2+\epsilon^2}$, the series converges uniformly on the integration interval $[a,b]$, allowing us to exchange the sum with the integral obtaining
\begin{equation}\label{eq:expgs}
    \left\langle \frac{1}{x - \eta + i \epsilon}, \phi(x) \right\rangle = 
    \sum\limits_{k=0}^\infty 
    \eta^k\left\langle \frac{1}{(x + i \epsilon)^{k+1}},\phi(x)\right\rangle\,.
\end{equation}
We thus obtain a convergent power series in $\eta$ without imposing any analyticity assumption on the test function $\phi$. This result is a special case of the standard theorem on parameter-dependent integrals: if the integrand is holomorphic in a parameter and suitable uniform integrability conditions are satisfied, then the integral is holomorphic in that parameter even if the integrand is not analytic in the integration variable.\footnote{More precisely, one may formulate the result in terms of locally uniform domination of the integrand, or equivalently use standard theorems on differentiation and integration of holomorphic parameter-dependent functions.}

In the limit $\epsilon\to0^+$ eq.\,(\ref{eq:expgs}) still holds with a finite convergence radius but only provided additional conditions, which we now discuss, are satisfied. One such condition is the analyticity of the test function at the singular point $x=0$. This is necessary because the imaginary part of the distribution is the delta function, which acting on the test function gives just $\phi(\eta)$. The convergence radius of the distribution acting on $\phi$ is thus equal to the radius of the convergence of the Taylor series for $\phi$ at $x=0$. We show below that analyticity at the singularity is also sufficient for the validity of eq.\,(\ref{eq:expgs}), with the additional requirement that the integration extremes $a$ and $b$ are away from the singularity.

We regard eq.\,(\ref{eq:distr_def11}) as an integral in the complex plane over the real interval $[a,b]$ and we suppose that $\phi$ admits a holomorphic extension to a neighbourhood of the origin, shown as a grey circle in figure\,\ref{fig:cp}. The integrand
\begin{equation}
\frac{\phi(z)}{z-\eta+i\epsilon}\,,
\end{equation}
is holomorphic in the same region, except at $z = \eta - i \epsilon$. The denominator is singular on the lower half-plane (since $\epsilon>0$) and for $\epsilon\to0^+$ and for sufficiently small (but finite) $\eta$ the pole is inside the analyticity domain of $\phi$. We can thus deform the portion of the integration contour near the origin into the upper half of the complex plane pushing it at a finite distance from the singularity. The contour can be left on the real axis outside the analyticity region as shown in the figure. No analyticity assumption on $\phi$ is thus required at $x\neq0$. This is important because in the integrals encountered in the main text there are (integrable) singularities in $\phi$ at the integration extremes.

\begin{figure}[t]
\centering
\begin{tikzpicture}[
    scale=1.0,
    >={Latex[length=2.2mm]},
    contour/.style={thick},
    defcontour/.style={gray!60, thick}
]

\def\a{-1.3}
\def\b{2.0}
\def\r{0.7}
\def\xp{0.35}   
\def\yp{-0.10}  

\pgfmathsetmacro{\xmA}{(\a-\r)/2}
\pgfmathsetmacro{\xmB}{0}
\pgfmathsetmacro{\xmC}{(\r+\b)/2}

\begin{scope}[xshift=0cm]

\tikzset{
  axis/.style={->, semithick, >={Latex[length=1.5mm]}}
}

\fill[gray!15] (0,0) circle (\r);

\draw[axis] (\a-0.55,0) -- (\b+0.55,0);
\draw[axis] (0,-0.8) -- (0,1.8);

\draw[contour] (\a,0) -- (\b,0);


\draw[defcontour] (-\r,0) arc[start angle=180,end angle=0,radius=\r];


\draw[semithick] (\xp-0.055,\yp-0.055) -- (\xp+0.055,\yp+0.055);
\draw[semithick] (\xp-0.055,\yp+0.055) -- (\xp+0.055,\yp-0.055);

\fill (\a,0) circle (1.2pt);
\fill (\b,0) circle (1.2pt);

\node[left] at (\a+0.1,0.2) {$a$};
\node[right] at (\b-0.1,0.25) {$b$};
\node[above right] at (0.3,0.45) {$\gamma$};


\end{scope}

\begin{scope}[xshift=6.6cm]

\tikzset{
  axis/.style={->, semithick, >={Latex[length=1.5mm]}}
}

\fill[gray!15] (-1.7,1.55) rectangle (2.3,-0.7);
\draw[axis] (\a-0.55,0) -- (\b+0.55,0);
\draw[axis] (0,-0.8) -- (0,1.8);

\draw[contour] (\a,0) -- (\b,0);



\draw[defcontour] (\a,0) arc[start angle=180,end angle=0,radius=-\a];


\draw[semithick] (\xp-0.055,\yp-0.055) -- (\xp+0.055,\yp+0.055);
\draw[semithick] (\xp-0.055,\yp+0.055) -- (\xp+0.055,\yp-0.055);

\fill (\a,0) circle (1.2pt);
\fill (\b,0) circle (1.2pt);

\node[left] at (\a+0.1,0.2) {$a$};
\node[right] at (\b-0.1,0.25) {$b$};
\node[above right] at (0.58,0.95) {$\gamma$};

\end{scope}

\end{tikzpicture}
\caption{{\bf{Left:}}
A contour deformation whose extension is limited by the analyticity domain of $\phi$, represented as a shaded region. {\bf{Right:}}
The contour with maximal distance from the origin, of radius $r=\min\{|a|,|b|\}$.
\label{fig:cp}
}
\end{figure}

On the deformed integration contour $\gamma$, the geometric series
\begin{equation}
\frac{\phi(z)}{z - \eta + i \epsilon} = \sum_{k=0}^{\infty}
\eta^k\frac{\phi(z)}{(z+i\epsilon)^{k+1}}\,,
\end{equation}
has a uniform convergence radius $\rho$ that is given by 
\begin{equation}
\rho=\min_{z\in\gamma}|z+i\epsilon|\simeq \min_{z\in\gamma}|z|\,.
\end{equation}
This is finite for $\epsilon\to0^+$ and it is equal to the minimal distance of the contour from the origin. For $|\eta|<\rho$, the series can thus be integrated term by term on the $\gamma$ contour. Furthermore all the integrals in the series can be deformed back to the real $[a,b]$ interval because their only pole is on the negative half-plane at $z=-i\epsilon$. We thus recover eq.\,(\ref{eq:expgs}), with a finite convergence radius $\rho$.

The left panel of figure\,\ref{fig:cp} shows a configuration where the convergence radius $\rho$ is limited by the extension of the analyticity region (depicted as a circle) of the test function. On the right panel instead the analyticity region is larger and it is not a limiting factor. The convergence radius is $\rho=\min\{|a|,|b|\}$ that is the maximum allowed by the fact that the contour must start and end on the boundaries of the interval.

The discussion so far implicitly assumes that $a<0$ and $b>0$ such that the integration interval contains the singular point. If this is not the case, no contour deformation is needed and the convergence radius is $\rho=a$ if $0<a<b$ or $\rho=-b$ if $a<b<0$. If instead the singularity coincides with one of the integration boundaries (i.e., $a=0$ or $b=0$) it cannot be avoided. Eq.\,(\ref{eq:expgs}) is violated in this case.

The series of distributions encountered in the expansion\,(\ref{eq:expgs})---and in the analogue formula with $-i\epsilon$ regulator---are derivatives of the $1/(x\pm i\epsilon)$ distribution
\al{\label{eq:1/x^n_expansion}
\frac1{(x\pm i\epsilon)^{n+1}} = &
\frac{(-1)^n}{n!}\frac{d^n\;}{dx^n}\frac1{x\pm i\epsilon}
=\frac{(-1)^n}{n!}{\rm{pv}}^{(n)}\big[1/x\big]
\mp i\pi\frac{(-1)^n}{n!}\delta^{(n)}(x)\\
=&
{\rm{Pf}}\big[1/x^{n+1}\big]
\mp i\pi\frac{(-)^n}{n!}\delta^{(n)}(x)
\,.\nonumber
}
In the equation, the Hadamard finite part of $1/x^{n+1}$ is defined as
\al{
 \left\langle {\rm Pf}\frac{1}{x^{n+1}}, \phi\right\rangle\hspace{-2pt} \equiv \hspace{-2pt}\lim_{\epsilon \to 0^+} \int_a^b \phi(x)\, {\rm Re}\frac{1}{(x \pm i \epsilon)^{n+1}} dx
 = \hspace{-2pt} \lim_{\epsilon \to 0^+} \frac{(-1)^{n}}{n!} \int_a^b \phi(x)\, {\rm Re} \frac{d^{n}}{dx^{n}}\frac{1}{x \pm i \epsilon} dx\,.\nonumber
}
Using the second expression and integrating by parts $n$ times we can rewrite the Hadamard finite part as a principal value integral of the $n$-th derivative of $\phi$ plus boundary terms. For $n=1$ 
\beq\label{eq:pf2}
\left\langle {\rm Pf}\frac{1}{x^{2}}, \phi\right\rangle=
\frac{\phi(a)}{a}-\frac{\phi(b)}{b}
+\left\langle {\rm pv}\frac{1}{x}, \phi^{(1)}\right\rangle.
\eeq

\bibliographystyle{JHEP}
\bibliography{biblio.bib}

\end{document}

%% file: Figures/fig1.tex
\begin{figure}[t]
\begin{minipage}[c]{0.6\textwidth}
\centering
\begin{tikzpicture}
\begin{feynman}
\node [] (v1) at (-2,-2);
\node [] (v2) at (2,-2);
\node [label=left:\(k_1\)] (vv1) at (-0.1,-0.2) ;
\node [label=right:\(k_1\)] (vv2) at (0.1,-0.2) ;
\draw (v1) -- (vv1)  
      node[pos=0.85, below=-1pt, sloped] {$\rightarrow$};
\draw (v2) --  (vv2)   
      node[pos=0.85, below=-1pt, sloped] {$\leftarrow$};
\node [label=left:\(k_2\)] (vvv1) at (-0.1,-1.1) ;
\node [label=right:\(k_2\)] (vvv2) at (0.1,-1.1) ;
\draw (v1) -- (vvv1)   
      node[pos=0.85, below=-1pt, sloped] {$\rightarrow$};
\draw (v2) --  (vvv2)   
      node[pos=0.85, below=-1pt, sloped] {$\leftarrow$};
\node (vvvv1) at (-0.1,-2) ;
\node (vvvv2) at (0.1,-2) ;
\draw[scalar] (v1) -- (vvvv1) ;
\draw[scalar] (v2) --  (vvvv2) ;
\node [] (vvvvv1) at (-0.1,-2.9) ;
\node [] (vvvvv2) at (0.1,-2.9) ;
\draw[scalar] (v1) -- (vvvvv1) ;
\draw[scalar] (v2) --  (vvvvv2) ;
\node [ label={[label distance=-10pt]above:\(\0{p}_1\)}] (w1) at (-3.5,-0.5);
\node [label={[label distance=-10pt]above:\(\1{p}_1\)}] (w2) at (3.5,-0.5);
\draw  (w1) --  (v1) 
      node[pos=0.1, above=-1pt, sloped] {$\rightarrow$};
\draw  (w2) --  (v2)  
      node[pos=0.1, above=-1pt, sloped] {$\leftarrow$};
\node [ label={[label distance=-10pt]left:\(\0{p}_2\)}] (z1) at (-4,-2);
\node [label={[label distance=-10pt]right:\(\1{p}_2\)}] (z2) at (4,-2);
\draw  (z1) --  (v1) 
      node[pos=0.1, above=-1pt, sloped] {$\rightarrow$};
\draw  (z2) --  (v2)  
      node[pos=0.1, above=-1pt, sloped] {$\leftarrow$};
\node [] (x1) at (-3.7,-3.0);
\node [] (x2) at (3.7,-3.0);
\draw[scalar]  (x1) --  (v1) ;
\draw[scalar]  (x2) --  (v2) ;
\filldraw[color=gray!50, fill=gray!5, very thick](-2,-2) ellipse (1.2 and 0.9);
\filldraw[color=gray!50, fill=gray!5, very thick](2,-2) ellipse (1.2 and 0.9);
\vertex (v) at (-2,-2) {${\mathcal{M}}\big(\{\0{{p}}\};\hspace{-1pt}\{{k}\}\big)$};
\vertex (v) at (2,-2) {${\mathcal{M}}^*\hspace{-1pt}\big(\{\1{{p}}\};\hspace{-1pt}\{{k}\}\big)$};
\end{feynman}
\end{tikzpicture}
\end{minipage}
\begin{minipage}[c]{0.4\textwidth}
\centering
\begin{align}
&p_1\equiv\frac{\0{p}_1+\1{p}_1}2\,,\quad
\di{p}_1\equiv\0{p}_1-\1{p}_1\,,
\nonumber\\
&p_2\equiv\frac{\0{p}_2+\1{p}_2}2\,,\quad
\di{p}_2\equiv\0{p}_2-\1{p}_2\,,
\nonumber\\
&\ldots\qquad\qquad\quad\ldots
\nonumber\\
&p_{\rm{tot}}\equiv p_1+p_2+\ldots=k_{\rm{tot}}\,, 
\nonumber\\
&\di{p}_1+\di{p}_2+\ldots=0\,.\nonumber
\end{align}
\end{minipage}
\caption{{\bf{Left:}} Pictorial representation of eq.\,(\ref{eq:Mas1}).
{\bf{Right:}} The definition of the mean momenta and of the momentum differences with the corresponding conservation equations.\label{diag}}
\end{figure}

%

%% file: Figures/fig2.tex
\begin{figure}[t]
{
\begin{minipage}[c]{1\textwidth}
\centering
{
\begin{tikzpicture}
\begin{feynman}
\node [] (InsIN) at (-6.0,1.);
\node [dot] (int1) at (-3.5,1.);
\draw[thick] (InsIN) -- (int1) 
      node[pos=0.5,above=-19pt] {$\begin{array}{c}
      \u,\,m_\u\\[-0pt]
      \0{p}_\u,\1{p}_\u
      \end{array}
      $}
      node[pos=0.1, above=-1pt] {$\rightarrow$};
\draw [] (int1) -- (-1.5,1.)
      node[pos=0.5, above=-19pt]
      {$\begin{array}{c}
      \D,\,\mu_\D\\[-0pt]
      K_\D
      \end{array}
      $}
      node[pos=0.9, above=-1pt] {$\rightarrow$};
\draw [] (int1) -- (-1.5,0.95);
\draw [] (int1) -- (-1.5,1.05);
\node [] (StIN) at (-6.0,-1);
\node [dot] (int2) at (-3.5,-1);
\draw[thick] (StIN) -- (int2) 
      node[pos=0.5, above=-19pt]
      {$\begin{array}{c}
      \s,\,m_\s\\[-0pt]
      \0{p}_\s,\1{p}_\s
      \end{array}
      $}
      node[pos=0.1, above=-1pt] {$\rightarrow$};
\draw [] (int2) -- (-1.5,-1.)
      node[pos=0.5, above=-19pt]
      {$\begin{array}{c}
      \P,\,\mu_\P\\[-0pt]
      K_\P
      \end{array}
      $}
      node[pos=0.9, above=-1pt] {$\rightarrow$};
\draw [] (int2) -- (-1.5,-0.95);
\draw [] (int2) -- (-1.5,-1.05);
\draw[thick] (int1) -- (int2) 
      node[pos=0.47, right=0pt] {$\begin{array}{c}
      \i,m_\i\\[-2pt] 
      \0{q},\1{q}\end{array}$}
      node[pos=0.5, below=-1pt,sloped] {$\rightarrow$};
\node at (-6.5,0) {$i{\mathcal{M}}_t=$};
\filldraw[color=gray!50, fill=gray!5, very thick](-3.5,1.) ellipse (0.4 and 0.4);
\vertex (v) at (-3.5,1.) {${\mathcal{M}}_\D$};
\filldraw[color=gray!50, fill=gray!5, very thick](-3.5,-1.) ellipse (0.4 and 0.4);
\vertex (v1) at (-3.5,-1.) {${\mathcal{M}}_\P$};
\end{feynman}
\end{tikzpicture}}
\end{minipage}
}
\caption{Diagrammatic contribution to the $\u$-$\s$ collision featuring the $t$-channel exchange of the intermediate particle $\i$.\label{diag1}}
\end{figure}